\documentclass[showpacs,twocolumn]{revtex4}
\usepackage{epstopdf}
\usepackage{color}
\usepackage{bm}
\usepackage{graphicx}
\usepackage{amsmath}
\usepackage{amsfonts}
\usepackage{amssymb}
\usepackage{bezier} 
\usepackage{color} 

\mathchardef\Gamma="7100

 \def\bq{{\bf q}}
\def\bB{{\bf B}}

\def\bE{{\bf E}}
\def\bx{{\bf x}}
\def\by{{\bf y}}
\def\bp{{\bf p}}

\def\bz{{\bf z}}
\def\bj{{\bf j}}

\def\bA{{\bf A}}
\def\bS{{\bf S}}
\def\bk{{\bf k}}

\def\cS{{\cal S}}
\def\cL{{\cal L}}

\def\cP{{\cal P}}
\def\cA{{\cal A}}
\def\cO{{\cal O}}

\def\cV{{\cal V}}
\def\cM{{\cal M}}

\def\Im{{\mbox{Im}}}
\def\Re{{\mbox{Re}}}
\def\Tr{{\mbox{Tr}}}

\def\ve{{\varepsilon}} 
\def\half{{1\over 2}}
\def\be{\begin{equation}}
\def\ee{\end{equation}}
\def\bea{\begin{eqnarray}}
\def\eea{\end{eqnarray}}
\def\bv{{\bf v}}

\def\tP{{\tilde{P}}}
\def\tQ{{\tilde{Q}}}

\begin{document}
\title{Equilibrium Formulae for Transverse Magneto-transport of  Strongly Correlated Metals}

\author{Assa Auerbach}
\affiliation{Physics Department, Technion, 32000 Haifa, Israel}

\date{\today }
\begin{abstract}
Exact formulas for the Hall coefficient [A. Auerbach, Phys. Rev. Lett. {\bf 121}, 066601 (2018)], modified Nernst coefficient, and thermal Hall coefficient of metals are derived from the Kubo formula.  
These coefficients depend exclusively on {\em equilibrium susceptibilities}, which are significantly easier to compute than conductivities.  
For weak isotropic scattering, Boltzmann theory is recovered.  For strong scattering, well controlled methods for thermodynamic functions are available.
As an example, the Hall sign reversals of lattice bosons near the Mott insulator phases are determined.
Appendices include mathematical supplements and instructions for
calculating the coefficients.
 \end{abstract}

\maketitle
\section{Introduction}
 Computation of 
transport coefficients of strongly correlated metals, 
is challenging even for minimal model Hamiltonians. 
DC conductivities are particularly costly, since they involve real-time correlations of large systems in the limit of long times. 

The  Hall coefficent  $R_H$ -- the magnetic field derivative of the transverse DC resistivity at low fields --  seems to be an interesting exception. 
For isotropic bands, Boltzmann equation relates $R_H$  to the inverse carrier density. For more realistic band structures, $R_H$ is related to the
Fermi surface curvature~\cite{jones-zener,ziman,Ong}.  Thus, at least for isotropic scattering~\cite{Comm-tau}, $R_H$ is insensitive to the scattering timescale and depends only on equilibrium coefficients. (Here,  {``equilibrium coefficients''}  are defined as {\em static} derivatives of the free energy, which do not involve time dependent correlators).

$R_H$ raises intriguing questions:  (i) Is the Hall coefficient {\em in general} an equilibrium  property, beyond the validity of Boltzmann theory? (ii) Is there an explicit formula which expresses $R_H$ in terms of static susceptibilities?
 (iii)   Are there other equilibrium formulas for magneto-transport coefficients of resistive metals~\cite{Comm-Other}?

These questions are particularly relevant to  ``bad metals'', where scattering rates exceed the Fermi energy~\cite{badmetals,lindner} and quasiparticles are not well-defined.  Bad metals are known to exhibit ``Hall anomalies'' -- poorly understood magnetic field,  temperature, and doping dependences of the Hall coefficient, including unexpected sign reversals.
Hall anomalies have been observed in strongly disordered films~\cite{smith,kapitulnik},  
resistive phases of unconventional superconductors~\cite{hagen,Taillefer}, strongly correlated metallic paramagnets~\cite{Lior}, and more. Resolving the
origin of the Hall anomalies has been hampered by the innaplicability of Boltzmann equation, and the formidable numerical challenges of DC conductivities.

 In a recent paper~\cite{Assa-PRL},  Questions (i) and (ii), have been answered by
the derivation of a formula for $R_H$, which depends solely on equilibrium susceptibilities.
The formula is applicable to general interacting and disordered Hamiltonians. The coefficients 
are amenable to well controlled numerical algorithms including:  high temperature series~\cite{domb}, variational wavefunctions~\cite{DMRG},
Quantum Monte Carlo simulations~\cite{prokofiev,assaad} (in imaginary time), and more.    Most importantly,
the Hall coefficient  does not depend on real-time DC conductivities, which inherently involve less controlled and much costlier computations~\cite{Comm-RT,White,MaxEnt,snir-QMC}.
  
 This paper reviews and expands the derivation of the Hall coefficient formula~\cite{Assa-PRL}. It also answers Question (iii) by deriving two additional  equilibrium formulas  for transverse magneto-transport coefficients.  It opens up the possibility for feasibly computing  magnetotransport coefficients for strongly correlated Hamiltonians.
 
 Three formulas are presented in this paper:
  \begin{enumerate}
\item The Hall coefficient is  
\be
R_H \equiv  \sigma_{xx}^{-2}~{d\sigma_{H}\over dB}\Big|_{B\!=\!0} ,
\label{HallC-def}
\ee
where $\sigma_{H}$  and $\sigma_{xx}$ are the Hall and longitudinal conductivities respectively, and $B$ 
is the perpendicular magnetic field. The formula is
 \bea
R_H &=& R_H^{(0)}+ R_H^{\rm corr} , \nonumber\\
R_H^{(0)} &=& -\Im  { \left(j^x| [M,j^y]\right) -  \left(j^y| [M,j^x]\right)  \over \hbar \cV \mu_0^2 } ,
\label{HallC}
\eea
where $M$ is the total magnetization operator, $j^\alpha, ~\alpha=x,y$ are the uniform $(\bq\!=\!0$) electric currents, and  $\cV$ is the system's volume in $d$ dimensions.
$(A|B)$ is a static mutual susceptibility of operators $A$ and $B$,
\be
(A|B) \equiv -\partial_{h_A} \partial_{h_B} \Tr \log e^{- (\beta H -h_A A -h_B B)} \Big|_{h_A,h_B\!=\!0},
\label{AB-susc}
\ee
where $H$ is the zero field Hamiltonian.
$\mu_0= (j^\alpha| j^\alpha)/\cV$,  is
the zeroth moment  of the conductivity (f-sum rule).  
The correction $R_H^{\rm corr}$ is  defined by Eq.~(\ref{HallN}) in Section  \ref{Sec:HallC}.
\item The 
{\em modified Nernst coefficient} is
\be
W={1 \over \sigma_{xx}\kappa_{xx} } { d\alpha_{xy}\over dB}\Big|_{B\!=\!0} ,
\label{W-def}
\ee
where $\kappa_{xx}$ is the  thermal conductivity, and $\alpha_{xy}$ is the transverse thermoelectric (TTE) coefficient~\cite{behnia-review}.  The formula is
\bea
 W &\equiv &  W^{(0)}+ W^{\rm corr} ,\nonumber\\
  W^{(0)}&\equiv&   {   1\over \hbar     \mu_0^Q \mu_0  } \left( (j_Q^x | [M, j^y])- ( j_Q^y |[M, j^x]) \right) , 
 \label{nu-star}
 \eea
where $j_Q^\alpha, \alpha=x,y$  are the thermal currents, 
and $\mu_Q= (j_Q^\alpha| j_Q^\alpha)/\cV$ is
the thermal sum rule~\cite{shastry-sumrules}. 
The correction $W^{\rm corr}$ is  defined in Eq.~(\ref{nu-star1}) in Section  \ref{Sec:Nernst}.

\item The {\em  thermal Hall coefficient} is  
\be
R_{TH}= { 1\over \kappa_{xx}^2} {d\kappa_{xy} \over dB}\Bigg|_{B\!=\!0} ,
\label{RTH-def}
\ee
where $\kappa_{xy}$ is the thermal Hall conductivity. The formula 
is
\bea
R_{TH} & =&   R_{TH}^{(0)}+ R_{TH}^{\rm corr} ,\nonumber\\
R_{TH}^{(0)}&\equiv&   {   T  \over \hbar   \cV ( \mu_0^Q )^2  } \left( \left(j_Q^x | [M , j_Q^y]\right)\!-\! \left( j_Q^y | [M , j^y_Q]\right) \right) .
\label{R-TH}
\eea
The correction $R_{TH}^{\rm corr}$ is  defined in Eq.~(\ref{R-TH1}) in Section  \ref{Sec:RTH}.

\end{enumerate}

The correction terms  $R_H^{\rm corr}$, $W^{\rm corr}$, and $R_{TH}^{\rm corr}$
are sums over rational functions of equilibrium susceptibilities of  local operators. These operators are constructed by multiple commutators of $M$, $H$ and the uniform electrical and thermal currents.

Here we are interested in strongly correlated metals which are not  amenable to perturbative expansions or to Boltzmann's transport theory.   
The derivation of Eqs. (\ref{HallC}), (\ref{nu-star}) and (\ref{R-TH}) starts with the many-body Kubo formula in the  Lehmann (eigenstate) representation. 
Numerical evaluation of this representation requires exponentially large memory cost.
{\em Bogoliubov operator hyperspace} and {\em Krylov operators} formulation~\cite{Bogoliubov,MerminWagner,Mori} provide a very useful framework for our derivations.

The reader may not be a-priori familiar with hyperspace terminology, which will be fully defined in the following sections.
We note that hyperspace has been extensively used to generate memory functions for transport theory~\cite{forster,zwanzig,wolfle}.

Bogoliubov hyperspace provides essential advantages:
\begin{itemize}
\item  Avoids the prohibitive cost of exact diagonalization required for the Lehmann representation of the Kubo formula.
\item  Charts a direct route to  continued fraction expansions of conductivities of strongly correlated metals~\cite{MHLee,viswanath,Sandvik,lindner,khait},
\item Enables a convenient framework for differentiating the conductivities with respect to magnetic field.
\end{itemize}
The latter advantage is a key ingredient in the proofs given below. 

Application of our formulas to  models of electrons and bosons is instructive. 
The zeroth  terms  $R^{(0)}_H$, $W^{(0)}$, and $R^{(0)}_{TH}$ recover Boltzmann equation result in the constant lifetime approximation. Anisotropic lifetime effects  appear in higher order corrections.
For  lattice bosons, we locate the Hall sign changes in the vicinity of the Mott insulator lobes.   
From these examples we learn that low energy renormalization of the microscopic Hamiltonian can greatly enhance the
 relative magnitudes of the zeroth terms relative to the harder-to-compute correction terms.

This paper is organized as follows. Section \ref{Sec:Kubo} introduces the Kubo formulas for the Hall  and TTE conductivities in
Bogoliubov hyperspace notations.  Section \ref{Sec:HallC} derives the Hall coefficient formula, Eq.~(\ref{HallN}).
Section \ref{Sec:Nernst} derives the modified Nernst
coefficient formula, Eq.~(\ref{nu-star1}). Section \ref{Sec:RTH} derives the thermal Hall coefficient formula, Eq.~(\ref{R-TH1}).
  
Section \ref{Sec:Appl} discusses applications of the formulas to effective Hamiltonians, band electrons, and strongly interacting lattice bosons. 

Section \ref{Sec:Other} is peripherally connected to the bulk of this paper.  From the Kubo formula,  some known relations between equilibrium observables and  conductivities are derived: 
The Streda formulas~\cite{Streda,McD}, Chern numbers~\cite{TKNN,yosi,huber}, and  Hall-pumped polarization~\cite{Chern-pump-theory,prelov-ladder,Chern-pump-exp}. The derivation clarifies why these relations 
are restricted to bulk-incompressible,  {\em non-dissipative} systems where  $\sigma_{xx}\!=\!0$.

The paper is concluded by a summary and proposals for applications of our formulas to interesting models.

The appendices contain instructive technical details for computing the formulas.
Appendix \ref{App:Krylov} constructs Krylov bases in the Bogoliubov hyperspace.
Appendix  \ref{App:CF}  expands the longitudinal conductivities
$\sigma_{xx}(\omega) and \kappa_{xx}(\omega)$ as continued fractions. 
Appendix \ref{App:mom-rec} explains how to compute the moments, recurrents and magnetization matrix elements as equilibrium coefficients.
Appendix \ref{App:VER} describes the variational extrapolation of recurrents scheme, which obtains 
dynamical response functions from a finite set of moments.  Appendix \ref{App:OD} calculates the  Liouvillian Green function 
and shows how the DC conductivities factor out of the magneto-transport coefficients. This is the key  result which proves that the coefficients are purely equilibrium quantities.

\section{Kubo formula in hypespace notations}
\label{Sec:Kubo}
DC conductivities  of metals are defined  (using an infinitesimal $\ve$ prescription) by the following order of limits
\be
\sigma_{\alpha\beta} \equiv    \lim_{\omega\to 0} \lim_{\bq \to 0} \lim_{\ve\to 0} \lim_{ \cV \to \infty}  \sigma_{\alpha\beta}(\bq,\omega;\cV,\ve) ,
\label{limits}
\ee
where the dynamical conductivities are given by the Kubo formula
 \bea
 \sigma_{\alpha\beta}(\bq,\omega)  &=& {\hbar\over  \cV    }\Im \sum_{n,m}  { (\rho_n\!-\!\rho_m )    \langle m| j^\alpha_{  \bq} |n\rangle \langle n|j^\beta_{ -\bq} |m\rangle   \over  (E_m\!-\!E_n)(E_m\!-\!E_n-\hbar\omega\!-\!i\ve)  } ,  \nonumber\\
 &=&   { \hbar \over  \cV}   \Im  \left( j_{\bq}^\alpha   \Big|  \left( {1\over \cL-\hbar\omega-i\ve} \right)   \Big| j_{\bq}^\beta  \right) .
  \label{Kubo}
\eea
$j^\alpha_\bq $ are the  spatial Fourier components of currents, and $\alpha$ denotes both the transported quantity (charge or heat) 
and the direction of the current $x$ or $y$.
$E_n$ and $|n\rangle$ are the eigenenergies and eigenstates of  the grand Hamiltonian, $H-\mu N$, respectively.   
$\rho_n = e^{-\beta E_n}/ \Tr e^{-\beta( H-\mu N)}$ are Boltzmann weights. 
Henceforth, we avoid the ``$\lim$'' symbols for the DC limit, remembering the order of limits in (\ref{limits}).

For pedagogical simplicity, we restrict ourselves to a uniform magnetic field $\bB=B \hat{\bz}$. For $B=0$,  all response functions (after disorder averaging) obey  C4m  symmetry (reflections  and rotations around $\hat{\bz}$).  
Hence $\sigma_{xx}=\sigma_{yy}$ and $\sigma_{xy} = -\sigma_{yx}\equiv \sigma_H$.

The DC limit of a metal requires large system sizes, since $ \cV^{-1/d} \le \omega/v \to 0$ for some finite velocity scale $v$.
Memory requirements blow up as $e^{\cV}$, which is prohibitively costly, even for minimal Hamiltonians of strongly correlated metals, such as the Hubbard, t-J, and Kondo lattice models.
Hyperspace formulation, in the second line of  Eq.~(\ref{Kubo}), avoids the eigenstate representation.

{\em Hyperspace notations:}  The set of operators  $\{ A \}$ in Schroedinger Hilbert space define hypestates $|A)$
with the inner product~\cite{Mori}
\be
(A|B) \equiv    \sum_{nm} {\rho_n-\rho_m\over E_m-E_n} \langle m |A^\dagger|n\rangle \langle n | B|m\rangle  
\label{product1}
\ee
$(A|B)$ depends on temperature and is physically an equilibrium susceptibility given by Eq.~(\ref{AB-susc}). It can also be written as an imaginary-time
correlation function, see Eq.~(\ref{product}). We denote a normalized hyperstate by an angular bracket $|A\rangle$. 

The  Liouvillian $\cL$ is a hermitian hyperoperator that acts on hyperstate $|A)$  by $\cL |A)  = \big|[H, A]\big)$. 
The  DC hyper-resolvent can be separated into
\be
\left({1\over \cL-i\ve}\right)  \equiv    \left({1\over \cL}\right)'+i \left({1\over \cL}\right)''  
\ee
where
\bea  
\left({1\over \cL}\right)' &=&    {\cL \over \cL^2 +\ve^2} , \nonumber\\
\left({1\over \cL}\right)'' &=& {\ve \over \cL^2 +\ve^2}  .
\label{1overL}
\eea
We shall find it useful to write the inner product, Eq. (\ref{product1}),  as a trace in Schroedinger space
\be
(A|B)= - \Tr \rho  \left[\left({1\over \cL}\right)' A^\dagger , B\right] .
\label{trace}
\ee

By Eq. (\ref{Kubo}), the  Hall conductivity is  given by the  off-diagonal matrix element in hyperspace
\be
 \sigma_{H} = { \hbar \over  \cV}    \Im  \left( j^x \Big|  \left( {1\over \cL}\right)'  \Big|j^y\right) .  
 \label{Kubo-xy}
\ee
C4m and time reversal symmetries ensure that $\sigma_{H}$ is antisymmetric in $x  {\rightarrow \atop \leftarrow} y$, and in $B\to -B$.

Similarly, the antisymmetrized TTE coefficient is given by
\bea
\alpha_{xy}  =  { \hbar \over  T\cV}  \cA_{xy}\Im  \left( j_Q^x \Big|  \left( {1\over \cL}\right)'   \Big|j^y\right) - {c\over T\cV} \langle M^{\rm orb}\rangle ,\nonumber\\
 \label{Kubo-alphaxy}
\eea
where  $j_Q^\alpha$ is the thermal current in the $\alpha$ direction, and $\cA_{xy}$ is the antisymmetrizer defined by $\cA_{xy} f(x,y) = \half( f(x,y)-f(y,x))$.

The orbital magnetization in the $\hat{\bz}$ direction is
\be
M^{\rm orb} ={q\over 2c} \sum_{i=1}^N \bx_i \times \bv_i\cdot \hat{\bz}
\label{M-orb} ,\ee
which must be included in Eq.~(\ref{Kubo-alphaxy}) in order to satisfy Onsager's time reversal relations~\cite{Cooper}.  $N$ is the number of particles with charge $q$, positions $\bx_i$ and velocities $\bv_i$. 

Finally, the thermal Hall conductivity is  given by~\cite{Cooper}
\bea
\kappa_{xy} &=&  { \hbar \over  T\cV}  \cA_{xy}\Im  \left( j_Q^x \Big|  \left( {1\over \cL}\right)'   \Big|j_Q^y\right)\nonumber\\
&&~~~~ -  {2\over T\cV} \langle M^{Q}\rangle ,
 \label{Kubo-kxy}
\eea
where 
\be
M^{Q} ={1\over 4 } \sum_{i=1}^N \bx_i \times  \{ \bv_i , h_i -\mu \}\cdot\hat{\bz}
\label{M-Q}
\ee
is the thermal magnetization.

In Appendix \ref{App:CF}, the continued fractions of longitudinal conductivities $\sigma_{xx}$ and $\kappa_{xx}$ are derived, and the algorithm to compute
their recurrents is reviewed.
However,  transverse coefficients $\sigma_H$, $\alpha_{xy}$, and $\kappa_{xy}$ are off-diagonal matrix elements of the hyper-resolvent 
and, therefore, are not readily expressed as computable continued fractions.

\section{Derivation of the Hall coefficient formula}
\label{Sec:HallC}
While Eq.~(\ref{HallC-def}) is a ratio of transverse and longitudinal Kubo formulas [see Eq. (\ref{Kubo})], we find that the expression simplifies considerably by taking the derivative of $\sigma_H$ with respect to magnetic field~\cite{Comm-derivation}.
We thus {\em assume}  differentiability of the transport coefficients at zero field  in the paramagnetic, dissipative phase:
\be
 \sigma_{xx}(B) =\sigma_{xx} + \cO(B^2),~~~\sigma_H \propto B + \cO(B^3) .
 \label{metal}
 \ee
The conditions in Eq. (\ref{metal}) preclude zero resistivity and  quantum Hall phases, which
are amenable  to the equilibrium relations of Section \ref{Sec:Other}.

Using Eq.~(\ref{trace}),  the Hall conductivity in Eq. (\ref{Kubo-xy}) is written as
\bea
\sigma_{H} = -{ \hbar \over  \cV}  \Im    \Tr \rho \left[ \left({1\over \cL}\right)' j^x,   \left({1\over \cL}\right)' j^y \right]  .
\label{Kubo-sxy-general}
\eea        

\begin{figure}[!t]
\begin{center}
\includegraphics[width=8.5cm,angle=0]{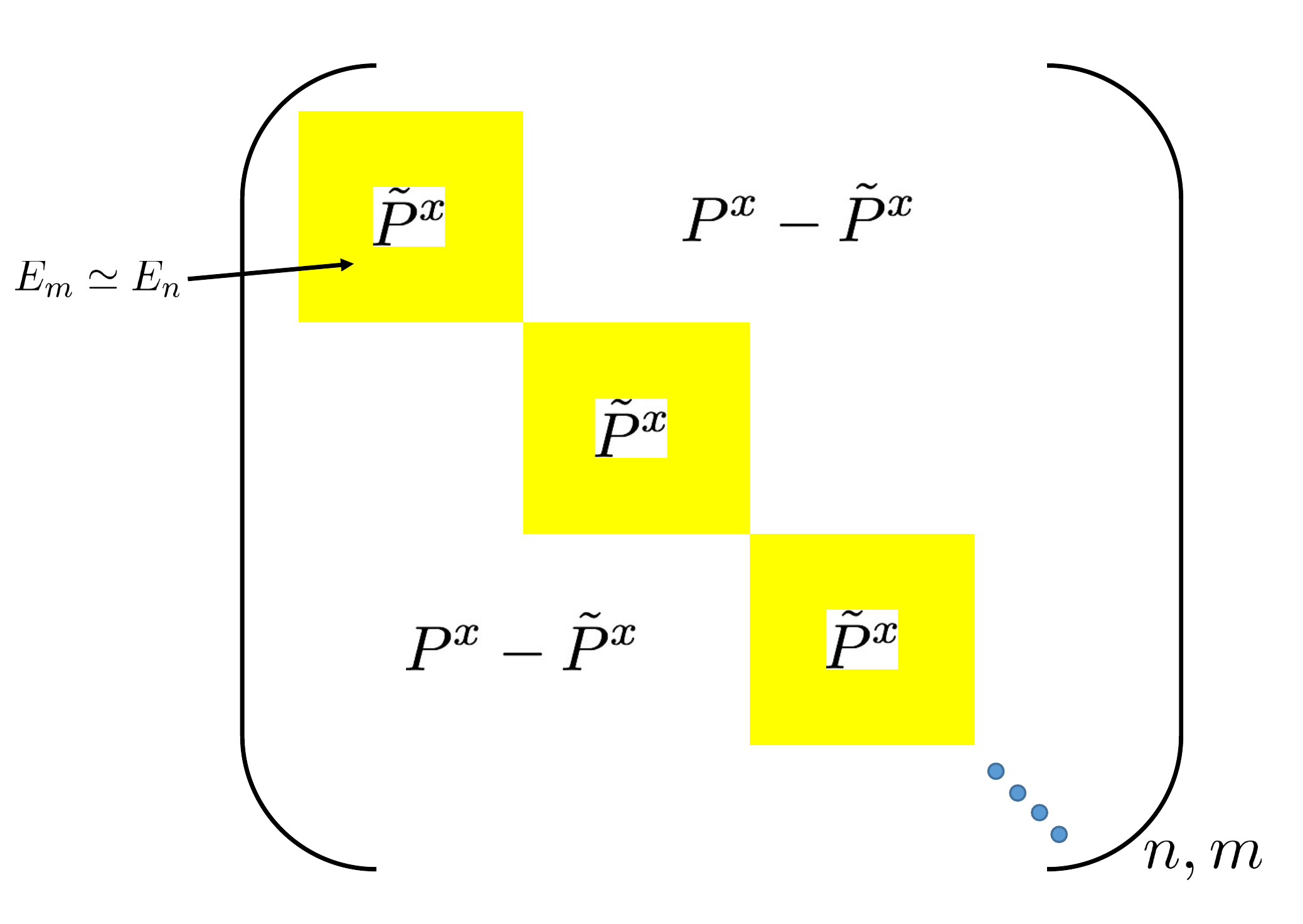}
\caption{The projection operator $P^x$ represented in the eigenenergy basis of $H$. The projected polarization $\tP^x$ in the degenerate subspaces
is marked by yellow blocks, and $P^x-\tP^x$ is supported in the white areas.
}
\label{Fig:P-tP}
\end{center}
\end{figure}

In non-periodic Euclidean space, one can define two commuting polarization  operators:
\be
P^\alpha  = q\sum_{i=1}^N x^\alpha_i , ~~~\alpha=x,y .
\label{pol}
\ee
The uniform electric currents are given by the operators
 \be
j^\alpha = {i \over \hbar }~ \cL P^\alpha  .
\label{jX}
\ee
Using Eq. (\ref{1overL}) in Eq. (\ref{jX}),  we obtain
\bea
\left({1\over \cL}\right)'   j^\alpha &=&{i\over \hbar}  \left({\cL\over \cL^2+\ve^2 }\right) \cL P^\alpha , \nonumber\\
&=& {i\over \hbar} \left( P^\alpha - \tP^\alpha\right),
\label{P-tP}
\eea
where $\tP^\alpha$ is the projection of  $P^\alpha$ onto the $\ve$-broadened kernel of $\cL$,
\be
\tP^\alpha \equiv \left({\ve^2 \over \cL^2+\ve^2 }\right) P^\alpha .
 \label{tP}
\ee
 In Fig.~\ref{Fig:P-tP}, the operators $P^x-\tP^x$ and $\tP^x$ are  depicted as submatrices of $P^x$ in the Lehmann representation.

Two points should be noted about $P^\alpha$:
(i) For systems with periodic boundary conditions in $\alpha$-direction (e.g. on a sphere, torus, cylinder, or ring),  
$P^\alpha and \tP^\alpha$ cannot be defined. For such cases, alternate expressions for $\left({1\over \cL}\right)' j^\alpha$ are given in Section \ref{Sec:Other}.
(ii)  For translationally invariant Hamiltonians (no spatially varying potentials), $q^{-1}\tP^\alpha = R^\alpha$ are 
the global guiding center symmetries of $H$. Their  algebra, $[R^x,R^y]= -i { \hbar c\over eB}$, gives rise to an extensive Landau-level degeneracy. In dissipative metals, which concern this paper, $R^\alpha$ are not  symmetries, and  Landau level degeneracy is lifted by potentials and interactions.

Finally, the Hall conductivity in Euclidean space can be written as
\bea
\sigma_{H} =  {1 \over \hbar \cV}  \Im    \Tr \rho \left[ P^x-\tP^x, P^y-\tP^y \right] , 
 \label{Kubo-sxy-comm}
\eea
where, by Eq. (\ref{limits}), we send $\ve\to 0$ after $\cV\to \infty$ to obtain the equilibrium conductivity.
The  ``bare''  electric polarizations in Eq. (\ref{pol}) are independent of $B$ and mutually commute: $[P^x,P^y]=0$. However, the contributions of $\tP^\alpha$ to the commutator in Eq. 
(\ref{Kubo-sxy-comm}) survive in the presence of a finite magnetic field, even as  the limit $\ve\to 0$ is taken, leading to $\sigma_H\ne 0$. This is shown by the 
expressions derived below.

Taking the derivative of  Eq.~(\ref{Kubo-sxy-comm}) with respect to magnetic field yields two terms
\bea
{d\sigma_H\over dB}\Bigg|_{B\!=\!0} &=& \Xi_\rho+ \Xi_\cM ,\nonumber\\
\Xi_\rho&=&-{q^2 \over\hbar \cV}  \Im\Tr {d\rho\over dB} \left[ P^x-\tP^x,P^y-\tP^y\right]_{B\!=\!0} ,\nonumber\\
\Xi_\cM&=&  - {  q^2 \over \hbar \cV}  \Im  \Tr \rho \left[ -{d\tP^x\over dB},P^y-\tP^y\right]_{B\!=\!0}   \nonumber\\
&& - {  q^2 \over \hbar \cV}  \Im  \Tr \rho \left[ P^x-\tP^x, -{d\tP^y\over dB }\right]_{B\!=\!0} .
\label{Xi-rho-M}
\eea
$\Xi_\rho$ is evaluated using the operator identity
\be
 {d\rho\over dB} =  \beta  \rho (M_d - \langle M \rangle)   - \left[  \rho, \left({1\over \cL}\right)' M \right],
 \ee
 where $M = -{\partial H \over \partial B}$ is the magnetization, and  $M_d$ is its energy-diagonal part $[H,M_d]=0$. Thus,
  \bea
 \lim_{B\to 0} \Xi_\rho &=&   {\beta\over \hbar \cV} \Im  \beta \left(  M  \Big|  \left[P^x\!-\!\tP^x,P^y\!-\!\tP^y \right] \right)      \nonumber\\
&&~~+       {1\over \hbar \cV} \Im  \Tr   \rho  (M_d \!-\! \langle M\rangle)   \left[ P^x\!-\!\tP^x,P^y\!-\!\tP^y\right]  \nonumber\\
&=& 0  .\label{Xirho}
\eea 
Both terms of $\Xi_\rho$ vanish  at zero magnetic field  by time reversal symmetry.

To evaluate $\Xi_\cM$, the derivative ${d \tP^\alpha \over dB} $ uses the hyperoperator identity
\be
{d\over dB} \left( {1\over \cO(B)}\right)   =  - {1\over \cO }  {d\cO\over dB}  {1\over \cO } , 
\ee
where $\cO = \ve^2/(\cL^2(B) + \ve^2)$. This yields
\bea
&& {d \tP^\alpha \over dB} =  - {\ve\over \cL^2 + \ve^2}(\cM \cL  +\cL \cM){\ve \over \cL^2 + \ve^2} P^\alpha , \nonumber\\
&&~~=  -    i\hbar   \left({1\over \cL}\right)'' \cM \left({1\over \cL}\right)''  j^\alpha  + \cL  \left({1\over \cL}\right)'' \cM \left({1\over \cL}\right)'' P^\alpha ,\nonumber\\
\label{dtXdB}
\eea
where $\cM = - {\partial \cL \over \partial B} \equiv [M ,\bullet ]$ is the  {\em hypermagnetization}.

Thus, casting $\Xi_\cM$ as an inner product using Eq. (\ref{trace}), and using the hermiticity of $\cL$, yields
\be
{d\sigma_H\over dB}\Bigg|_{B\!=\!0}  = -{ 2 \hbar \over \cV}\cA_{xy}  \Im\left( j^x \Big| \left({1\over \cL}\right)'' \cM \left({1\over \cL}\right)'' \Big| j^y\right) +\Xi' .
\label{dsxydB}
 \ee
 The second term vanishes
 \be
 \Xi' =  -{  \hbar \over \cV}\cA_{xy}  \Im\left(  \cL  j^x \Big|  \left({1\over \cL}\right)'' \cM \left({1\over \cL}\right)'' \Big| P^y\right) =0 ,
 \label{Xi'}
 \ee
due to the hermiticity of $\cL$ and the identity proven  in Appendix \ref{App:OD},
\be
\left({1\over \cL}\right)''~ \cL~ j^\alpha =0~ .
\label{delta-cl}
\ee

Now we simplify Eq.~(\ref{dsxydB}) by inserting resolutions of identities between the hyperoperators.
For that purpose, we introduce the Krylov basis of orthonormal  operators $\{  |n/_{j^\alpha}\rangle \}$, 
which  are constructed by sequentially applying $\cL$  to the root state, the current  $|j^\alpha )$,
and orthonormalizing. Details are provided in Appendix \ref{App:Krylov}.
We note that $\langle n/_{j^x}|m/_{j^y}\rangle=0$ due to the C4m symmetry at zero magnetic field.
 
 The Krylov bases provide partial resolutions of identity [see Eq. (\ref{ROI})]: 
\be
\sum_{n=0}^\infty  |n/_{j^\alpha}\rangle \langle |n/_{j^\alpha}  | =1_{\cS_{j^\alpha}} ,~~~\alpha=x,y ,
\label{ROI1}
\ee
where $\cS_{j^\alpha}$ is the  subspace  spanned by $ \{ \cL^{n}|j^\alpha) \}_{n=0}^\infty$.
 
Application of Eq. (\ref{ROI1}) on the respective sides of  $\cM$ in Eq.~(\ref{dsxydB}) yields a double sum
\bea
&&{d\sigma_H\over dB}\Bigg|_{B\!=\!0} =  -{ 2 \hbar \mu_0 } \cA_{xy}  \sum_{m,n=0}^\infty  \langle 0/_{j^x}\Big|\left({1\over \cL}\right)''\Big| m/_{j^x}\rangle \nonumber\\
 &&~~~~\times  \Im \langle m/_{j^x}| \cM | n/_{j^y}\rangle\langle n/_{j^y}\Big|\left({1\over \cL}\right)''\Big| 0/_{j^y}\rangle ,\nonumber\\
 &&= {  \hbar \mu_0 }  \sum_{m,n=0}^\infty G''_{0,m}G''_{n,0}  M''_{m,n} ,\nonumber\\
 && M''_{m,n}  \equiv    \Im\left(  \langle m/_{j^x}| \cM | n/_{j^y}\rangle - \langle m/_{j^y}| \cM | n/_{j^x}\rangle \right) .
    \label{dsdB}
 \eea
The imaginary hyperresolvent matrix lements $G''_{0,n}=G''_{n,0}$ are evaluated in Appendix \ref{App:OD} [see Eq.~(\ref{Gn0})]: 
\bea
G''_{n,0}&=&  \langle n/_{j^y}\Big|\left({1\over \cL}\right)''\Big| 0/_{j^y}\rangle\times  \delta_{n,{\rm even}} , \nonumber\\
G''_{2k,0} (0) &=&   - \sigma_{xx} ~  {R_k  \over \hbar \mu_0} , \nonumber\\
R_k &=& \prod_{j=1}^{k} \left( - { \Delta_{2j-1}\over \Delta_{2j} } \right) ,
\label{Rk}
\eea
which shows that the longitudinal conductivity $\sigma^2_{xx}$ factors out of the double sum in Eq. (\ref{dsdB}). Using the definition of  the Hall coefficient in Eq.~(\ref{HallC-def}),
$\sigma_{xx}^2$ cancels out from $R_H$.
{\em This is a  key result of the  derivation!} 

The final  formula for the Hall coefficient is thus 
 \bea
R_H &=&  R_H^{(0)} + R_H^{\rm corr} , \nonumber\\
R_H^{(0)} &\equiv& - \Im  { \left(j^x|\cM|j^y\right) -  \left(j^y|\cM|j^x\right) \over \cV \hbar \mu_0^2 } , \nonumber\\
R_H^{\rm corr} &\equiv& - {1\over \hbar \mu_0}  \sum_{i,k=0}^\infty (1-\delta_{i ,0}\delta_{k,0}) R_i R_k M''_{2i,2k} . \nonumber\\
\label{HallN}
\eea
Eq.~(\ref{HallN}) defines $R_H^{\rm corr}$, which was presented earlier in Eq.~(\ref{HallC}).
 $R_k$, defined by Eq. (\ref{Rk}),  depends on a finite set of conductivity recurrents $\Delta_i, i\le 2k$, as defined in Appendix \ref{App:CF}. A recipe for their computation is given in Appendix \ref{App:mom-rec}.
The hypermagnetization matrix elements  $M''_{2i,2k}$  require computing mutual susceptibilities of operators, such as $  \Big| 2i/_{j^x}\Big\rangle $ and $\Big| \left[ M, | 2k/_{j^y}\rangle \right] \Big)$. 

In a non-critical, paramagnetic  metal,  $R_H<\infty$. Therefore, the double sum $\sum_{i,k}$ in 
$R_H^{\rm corr}$ is expected to (conditionally) converge, and its terms to decrease as $i,k\to \infty$.  
The rate of convergence depends on the particular Hamiltonian, but it could be estimated by computing a finite sequence of terms. 

As shown  in Section \ref{Sec:Appl}, the relative magnitudes $R_H^{\rm corr}/R_H^{(0)}$ could be greatly decreased  
at low temperatures by renormalizing the microscopic Hamiltonian onto an effective Hamiltonian.

\section{The modified Nernst coefficient}
\label{Sec:Nernst}
In this section and the next (Section \ref{Sec:RTH}), the derivations follow similar steps as in the previous section. Hence the discussion is briefer. 
To define the thermal current we need to be more specific about the Hamiltonian $H$. 
 We consider $N$ particles (either bosons or fermions) of charge $q$ described by a general continuum Hamiltonian
\bea
H&=& \sum_{i=1}^N h_i , \nonumber\\
h_i &\equiv& h_1\left(\bp_i,\bx_i,\bS_i; B\right)  + {1\over 2} \sum_{j,j\ne i} U_{ij} ~.
\eea
Here, the single particle Hamiltonian $h_1$ includes  kinetic, potential, and  spin energies.
$U_{ij} $ is a short range, two-body  interaction.

In close analogy to the electric polarizations $P^\alpha$, we define the {\em thermal polarizations} 
\be
Q^\alpha  = \half \sum_{i=1}^N \{ x^\alpha_i,h_i-\mu\} ,~~~\alpha=x,y .
\label{XQ}
\ee

The  heat current is simply the time derivative of the thermal polarization
\be
  j^\alpha_Q (\bq\!=\!0)  \equiv {i\over \hbar}\cL Q^x  \simeq \half \sum_{i=1}^N \{ v^\alpha_i,h_i-\mu\} .
\label{jQ-Qx}
\ee
Henceforth we neglect, for notational simplicity, the nonlocal contributions to $j^\alpha_Q$ of the form  $(\bv_i+\bv_j)\cdot \nabla U_{ij}  (x^\alpha_i - x_j^\alpha )$~\cite{Hardy}.
These can be included, but they contribute minor effects for short  range interactions $U_{ij}$.

Following the  analogous derivation which led to Eq.~(\ref{Kubo-sxy-comm}),  we use Eqs.~(\ref{trace}) and (\ref{jQ-Qx}) to express Eq. (\ref{Kubo-alphaxy}) as
\be
\alpha_{xy}  =   {q \over  \hbar T\cV}   \cA_{xy} \Im \Tr \rho \left[ Q^x-\tQ^x,P^y-\tP^y \right] - {c\over T\cV} \langle M^{\rm orb}\rangle,
\label{alphaxy-comm}
\ee
where
\be
\tQ^x \equiv  \left({\ve^2 \over \cL^2+\ve^2 }\right) Q^x .
\ee

The commutator between thermal and electric polarizations is nonzero:
\be
\cA_{xy} [Q^x,P^y] = { i \hbar\over 2} \sum_i (x_i v^y_i - y_i v^x_i)= i {\hbar c \over q} M^{\rm orb} ,
\label{com1}
\ee 
which precisely cancels against the orbital magnetization term in Eq. (\ref{alphaxy-comm}), leaving us with
\be
\alpha_{xy}=    {q \over \hbar T V} \cA_{xy} \Im  \Tr \rho \left[ -\tQ^x,  P^y-\tP^y\right] + \left[ Q^x-\tQ^x, -\tP^y\right]  .
\label{alphaxy-ready}
\ee

Differentiating Eq. (\ref{alphaxy-ready}) with respect to $B$ at $B=0$ yields the following terms
\be
{d \alpha_{xy} \over dB}\Bigg|_{B\!=\!0} =   - { \hbar \over T V} \cA_{xy} \Im  \Tr \rho \left[-{ d\tQ^x\over dB} ,  P^y  \right] + \left[  P^x  ,  -{d\tQ^y \over dB} \right] ,
\label{axy-comm}
\ee
where we discard, as in Eq. (\ref{Xirho}),  time reversal symmetry breaking terms  from ${d\rho\over dB}$, as well as the (undifferentiated) $\tP,\tQ$ operators, which contribute corrections of $\cO(\ve)$.

$d\tQ^x/dB$ yields two terms
\bea
  {d \tQ^\alpha \over dB} &=&  -i \hbar  \left({1\over \cL}\right)'' \cM \left({1\over \cL}\right)'' j_Q^\alpha   \nonumber\\
&&  -  {\ve^2 \over \cL^2 + \ve^2} \sum_{i=1}^N x^\alpha_i m_i ,
\label{dtXQ}
\eea
where the second term contributes $\cO(\ve)$ to  Eq.~(\ref{axy-comm}) and can be discarded.

Following the analogous derivation of Eq.~(\ref{dsxydB}) leads to
\bea
&&{d\alpha_{xy} \over dB}  =   -{ 2 \hbar \over T\cV}\nonumber\\
&&~~~\times \cA_{xy}  \Im\left( j_Q^x\Big| \left({1\over \cL}\right)'' \cM \left({1\over \cL}\right)'' \Big| j^y\right) .
\label{alphaxy-cM}
\eea

The Krylov  {\em thermal } resolution  of identity is
\be
\sum_{n=0}^\infty |n/_{j_Q^\alpha}\rangle   \langle n/_{j_Q^\alpha}| = 1_{\cS_{j_Q^\alpha} } .
\label{ROI-T}
\ee
Inserting the thermal resolution of identity from Eq. (\ref{ROI-T}) on the left of $\cM$ in Eq.~(\ref{alphaxy-cM})  and the electric resolution of identity from Eq. (\ref{ROI1}) on its right results in 
\bea
 {d\alpha_{xy} \over dB}\Big|_{B\!=\!0} &=&  -{    \sigma_{xx} \kappa_{xx}\over \hbar    (\mu_0^Q \mu_0)^{1\over 2} }\nonumber\\
&& \times \sum_{i,k} R^Q_i R_k   {M^Q_{2i,2k}}'' ,\nonumber\\
  {M^Q_{2i,2k}}''&\equiv & \Im \left( \langle 2i/_{j_Q^x} | \cM | 2k/_{j^y}\rangle -  \langle 2i/_{j_Q^y} | \cM | 2k/_{j^x}\rangle\right) , \nonumber\\
\eea
where  $\mu_Q = {1\over  \cV} (j_Q^x|j_Q^x)$ is the thermal conductivity sum rule~\cite{shastry-sumrules}.

The factors
\be
R^Q_i= \prod_{j=1}^{i} \left( - { \Delta^Q_{2j-1}\over \Delta^Q_{2j} } \right) ,
\label{R-Q}
\ee
depend on the recurrents $\Delta^Q_n$ of the thermal conductivity $\kappa_{xx}$, as defined in Eq.~(\ref{CF-sigma}).

The  modified Nernst coefficient defined by Eq.~(\ref{W-def}) is
given by the formula
\bea
W & =&    {1 \over \sigma_{xx}\kappa_{xx} } { d\alpha_{xy}\over dB}\Big|_{B\!=\!0} ,\nonumber\\
&=& W^{(0)}+ W^{\rm corr} ,\nonumber\\
W^{(0)}&\equiv&   {   1\over \hbar     \mu_0^Q \mu_0  } \left( (j_Q^x | \cM | j^y)- ( j_Q^y \cM |j^x) \right) ,\nonumber\\
W^{\rm corr}&\equiv&    {   1\over \hbar   (\mu_0^Q \mu_0 )^{1\over 2} } \sum_{i,k} R^Q_i R_k   {M^Q_{2i,2k}}'' (1-\delta_{ i,0}\delta_{k,0}) .\nonumber\\
 \label{nu-star1}
\eea
This equation defines $W^{\rm corr}$, which was presented in Eq.~(\ref{nu-star}). 

 $W$ is related to the  Nernst coefficient $\nu$ as follows:
\bea
\nu &=&  {d\over dB} \left( {E_x \over - {dT\over dy}}\right)_{B\!=\!0} , \nonumber\\
 &=&  \left( \sigma_{xx}^{-1} {d\alpha_{xy}\over dB}  - R_H \alpha_{xx}\right)_{B\!=\!0} ,\nonumber\\
W &=&   {\nu +R_H \alpha_{xx} \over \kappa_{xx}} .
\eea
For special  particle-hole symmetric systems,  $R_H,\alpha_{xx}=0$, and the relation simplifies to  $W= \nu/\kappa_{xx}$.

\section{The  Thermal Hall Coefficient}
\label{Sec:RTH}
The thermal Hall coefficient is the derivative of the  thermal Hall {\em resistivity} with respect to magnetic field at zero field.
The  thermal Hall conductivity in Eq.~(\ref{Kubo-kxy}) is given in Euclidean geometry by
\be
\kappa_{xy}  =   {1 \over  \hbar T\cV}   \cA_{xy} \Im \Tr \rho \left[ Q^x-\tQ^x,Q^y-\tQ^y \right] - {2\over T\cV} \langle M^Q\rangle ,
\label{kxy-comm}
\ee
where the thermal magnetization correction $M^Q$ is  defined by Eq.~(\ref{M-Q}).

The antisymmetrized commutator between ``bare''  thermal polarizations  yields
\bea
\cA_{xy} [Q^x,Q^y] &=& { i \hbar\over 2} \sum_i \left(x_i \{ v^y_i , h_i-\mu\} - y_i  \{ v^x_i , h_i-\mu\} \right) , \nonumber\\
&= &i {2 \hbar  } M^Q ,
\label{com2}
\eea
which precisely cancels against the second term in Eq. (\ref{kxy-comm}), leaving us with
\be
\kappa_{xy}=    {1\over \hbar T V} \cA_{xy} \Im  \Tr \rho \left[ -\tQ^x,  Q^y-\tQ^y\right] + \left[ Q^x-\tQ^x, -\tQ^y\right]  .
\label{kxy-ready}
\ee

Differentiating $\tQ^\alpha$ with respect to $B$, and discarding all terms of order $\ve$ yields
\bea
&&{d\kappa_{xy} \over dB}  =   - { 2 \hbar \over T\cV}\nonumber\\
&&~~~\times \cA_{xy}  \Im\left( j_Q^x\Big| \left({1\over \cL}\right)'' \cM \left({1\over \cL}\right)'' \Big| j_Q^y\right) .
\label{kxy-cM}
\eea

Now we insert two thermal resolutions of identities from Eq. (\ref{ROI-T}) and divide out $\kappa_{xx}^2$, as defined by Eq.~(\ref{RTH-def}), to obtain 
\bea
R_{TH} & =&   R_{TH}^{(0)}+ R_{TH}^{\rm corr},\nonumber\\
R_{TH}^{(0)}&\equiv&   {   T  \over \hbar    ( \mu_0^Q )^2  } \left( \left(j_Q^x | \cM | j_Q^y\right)- \left( j_Q^y |\cM |j^x_Q\right) \right),\nonumber\\
R_{TH}^{\rm corr}&\equiv&    {   T  \over \hbar    \mu_0^Q } \sum_{i,k} R^Q_i R^Q_k  {M^{QQ}_{2i,2k}}''(1-\delta_{ i,0}\delta_{k,0}),\nonumber\\
 {M^{QQ}_{2i,2k}}'' &\equiv & \Im \left(  \langle 2i /_{j_Q^x} | \cM | 2k/_{j_Q^y}\rangle -  \langle 2i /_{j_Q^y} | \cM | 2k/_{j_Q^x}\rangle \right) . \nonumber\\
\label{R-TH1}
\eea
The factors $R^Q_n$ are defined in Eq. (\ref{R-Q}). 
This equation defines $R_{TH}^{\rm corr}$, which was presented in Eq.~(\ref{R-TH}). 

\section{Applications to Effective Models}
\label{Sec:Appl}
It is greatly advantageous  at low temperatures to replace the microscopic Hamiltonian $H(\bA)$, where $\bA$ is the electromagnetic vector potential,
 by an effective Hamiltonian $\bar{H}(\bA)$ for two
reasons: 
\begin{enumerate}
\item  Reduction of the Hilbert space size, which greatly facilitates numerical computations.
\item  Rearrangement of the sums in Eqs. (\ref{HallN}) and (\ref{nu-star1})  by increasing the relative size of $R_H^{(0)}$ relative to  $R_H^{\rm corr}$.
\end{enumerate}
Eqs. (\ref{dsxydB}), (\ref{alphaxy-cM}), and (\ref{kxy-cM}) show that the two coefficients, at low temperatures,  are determined solely by the low energy part of Hilbert space. 
Let us examine Eq.~(\ref{dsxydB})  in the Lehmann representation 
\bea
&&{d\sigma_H\over dB}\Bigg|_{B\!=\!0}  = -{ 2 \pi^2 \hbar \over \cV}\cA_{xy}  \sum_{nmk} {\rho_m\!-\!\rho_n\over 
E_n\!-\!E_m } j^x_{nm} \delta(E_m\!-\!E_n) \nonumber\\
&&~~\times \left( M_{mk}  j^y_{kn} \delta(E_k\!-\!E_n) - j^y_{mk} M_{kn} \delta(E_m\!-\!E_k)  \right)) . 
\eea
The  energy conserving $\delta$ functions ensure that all participating states in the sum are (up to order $\ve$) degenerate $E_n \simeq E_m \simeq  E_k$ and restricted by  
Boltzmann weights to energies less than  some cut-off $\Lambda >   k_BT$.
We can therefore substitute  $H\to\bar{H}$, which shares the same low energy spectrum in a reduced Hilbert space, i.e.
\be
\bar{E}_n = E_n, ~~~E_n \le \Lambda .
\ee
All currents and magnetization in Eq. (\ref{dsxydB}) should also be replaced by their renormalized counterparts  given by
\bea
j^\alpha\to \bar{j}^\alpha &=&  -c {\partial \bar{H} \over \partial A^\alpha} ,\nonumber\\
M\to \bar{M} &=&  -  {\partial \bar{H} \over \partial B} .
\eea
Each individual term in the summation formulas  
is altered  by the renormalization,  since the Krylov bases, recurrents, and hypermagnetization matrix elements all depend on the renormalized operators. However,
an {\em exact renormalization}  must  leave Eq.~(\ref{dsxydB}) identical to that of the microscopic Hamiltonian. 

In many practical circumstances,  approximate renormalization are implemented. These include 
Schrieffer-Wolff transformations~\cite{SW}, Brillouin-Wigner perturbation theory~\cite{IEQM},  and  Contractor Renormalization (CORE)~\cite{CORE-marvin,CORE-altman,p6}. 

As a demonstration of the advantages of effective Hamiltonians, we compute $R_H^{(0)}$ for
a microscopic Hamiltonian
\be
H= \sum_{i=1}^N  { (\bp- {q\over c} \bA)^2\over 2m } + V(\bx_i) +\half  \sum_{i\ne j} U(\bx_i-\bx_j) +V^{\rm dis} ,
\ee
where $V$ is a periodic lattice potential, and $V^{\rm dis}$ describes a disorder potential.
The microscopic currents and magnetization obey
\bea
j^\alpha &=& q \sum_i { \bp^\alpha \over m } ,\nonumber\\
M &=& {q \over 2m c} \sum_i \bx_i \times  \bp_i  , \nonumber\\
\left[ M, j^\alpha \right] &=& {iq\hbar \over 2 m c}  \sum_\beta   \epsilon_{\alpha\beta} j^\beta , \nonumber\\
\Im  ( j^\alpha | \cM | j^\beta) &=&{  \hbar  V \over 2c} \mu_0  \epsilon_{\alpha\beta} , \nonumber\\
\mu_0 &=& {1\over \hbar \cV} \Im  \langle \left[ P^x,q\sum_i { p^x_i\over m} \right] \rangle , \nonumber\\
&=& {N q^2 \over   \cV m } ,
\label{micro-op}
\eea
where $\epsilon_{\alpha\beta}$ is the antisymmetric tensor.

By Eq. (\ref{micro-op}), the zeroth Hall coefficient term is inversely proportional to the total density
\be
R^{(0)}_H = {\cV\over N q c} .
\label{RH-micro}
\ee

\subsection{A single conduction band}
If the chemical potential lies within a single band,  separated by a large interband gap from other bands,
it is possible to describe the low spectrum by an effective
single band model
\be
\bar{H}= \sum_{\bk s} ( \epsilon_{ks} -\mu) c^\dagger_{\bk s} c_{\bk s} +\sum_{\bk s} \bar{V}^{\rm dis}_{\bk\bk'} c^\dagger_{\bk,s} c_{\bk' s} ,
\label{H-SB}
\ee
where $c^\dagger_{\bk,s}$ creates a band electron of charge $e$ and spin $s$ at lattice wavevector $\bk$. $\epsilon_\bk$ is the band dispersion, and 
$\bar{V}^{\rm dis}$ is the intraband disorder potential.
   
\begin{figure}[!t]
\begin{center}
\includegraphics[width=7.5cm,angle=0]{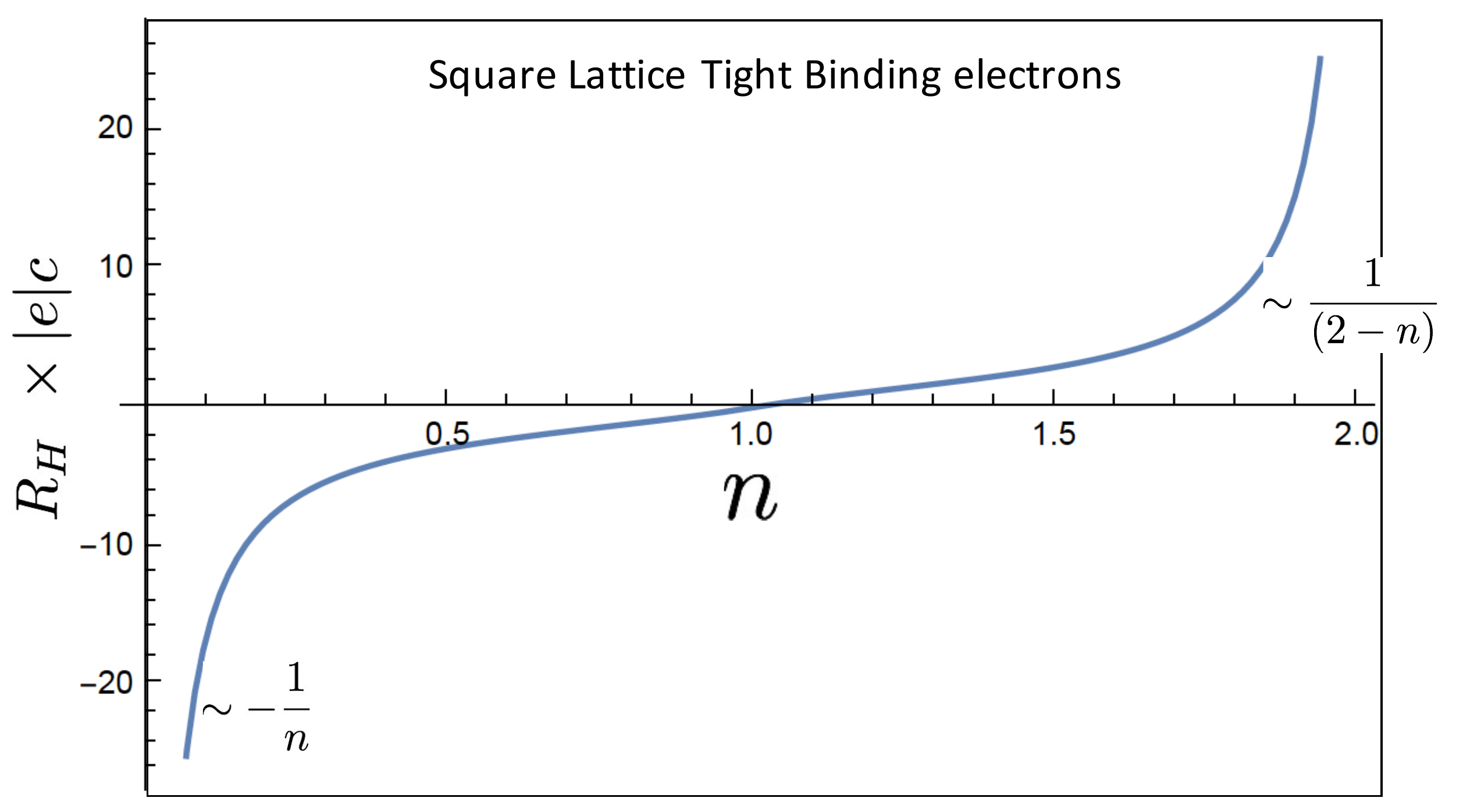}
\caption{Hall coefficient  versus electron filling, at low temperature, for the weakly disordered square lattice tight binding model, as given by Eq.~(\ref{BS1}).  }
\label{fig:SLTB}
\end{center}
\end{figure}

The single-band currents and magnetization are
\bea
\bar{j}^\alpha &=&  e \sum_{ks} v^\alpha_{\bk s} c^\dagger_{\bk s} c_{\bk s} , \nonumber\\
\bar{M}& =&  {i e\hbar\over  2c}  \sum_{ks}  c^\dagger_{\bk s} \left(v^y_{\bk s}  { d\over d k^x }-  v^x_{\bk s} { d\over d k^y } \right)c_{\bk s} ,
\eea
where $v_\bk^\alpha = {\partial \epsilon_\bk \over \partial k^\alpha}$.
Hence
\bea
\bar{R}_H^{(0)} &=& - {1\over   \cV} { \Im \left( \left(\bar{j}^x|\bar{\cM}|\bar{j}^y\right) - \left(\bar{j}^y|\bar{\cM}|\bar{j}^x\right) \right) \over \hbar \mu_0^2 } ,\nonumber\\
&=&  2 {e^3   \over  c \mu_0^2 } \int {d^d k \over (2\pi)^d} \left(-{\partial f \over \partial \epsilon}\right) F_\bk ,
\label{BS1}
\eea
where the mean Fermi surface curvature is given by
\be 
F_\bk =  (v^x_\bk)^2  { \partial^2 \epsilon_\bk \over (\partial k^y)^2} +   (v^y_\bk)^2  { \partial^2 \epsilon_\bk \over (\partial k^x)^2}  -  2 v^x_\bk v^y_\bk  { \partial^2 \epsilon_\bk \over \partial k^y \partial k^x}  ,
\ee
and the zeroth moment (f-sum rule) is
\be
\mu_0 = 2 e^2  \int {d^d k \over (2\pi)^d} \left(-{\partial f \over \partial \epsilon}\right)  |v^x_\bk  |^2 .
\ee
Eq.~(\ref{BS1}) recovers  Boltzmann equation result~\cite{jones-zener,ziman} in the case of wavevector-independent scattering time. In Fig.~\ref{fig:SLTB}, Eq.(\ref{BS1}) for  the square lattice tight  model is plotted as a function of electron filling.

A single parabolic bandstructure 
\be
\epsilon_\bk = -\mbox{\rm sign}(e ) { \hbar^2 |\bk|^2\over 2m^*},
\label{parabolic}
\ee
 where $e<0$ ($e>0$) describes  electrons (holes) respectively, yields
$\mu_0 = {ne^2 \over m^*}$.  Its Hall coefficient is equal to the famous Drude  result
\be
\bar{R}_H^{(0)}={1\over n e  c},
\label{drude}
\ee
where $n e$ is the charge density of this single band.

The corrections  $\bar{R}_H^{\rm corr}$ depend only on the weak impurity scattering $\bar{V}^{\rm dis}$, since
\be
\bar{\cL} j^y = e \sum_{\bk,\bk'} \bar{V}^{\rm dis}_{\bk,\bk'} (v^x_\bk - v^x_{\bk'}) c^\dagger_{\bk s}  c_{\bk' s} .
\label{Corr-Boltz}
\ee
Thus the factor $\Delta_1 / \Delta_2$ which enters the coefficients $R_k, k\ge 1 $, is suppressed as $\cO(\sqrt{ \langle ( \bar{V}^{\rm dis})^2 \rangle}/\epsilon_F \ll 1$ at weak disorder, where $\epsilon_F$ is the Fermi energy.

Recall that the  $R_H^{(0)}$ of  Eq.~(\ref{RH-micro}) was inversely proportional to the {\em total} density $N/\cV$,
including all core and  valence electrons.   For the effective single band model, the corrections were found to be suppressed at weak disorder,  $\bar{R}_H^{\rm corr}/\bar{R}_H^{(0)} \ll 1$.
Therefore, for the original microscopic Hamiltonian, $R_H^{\rm corr} $ is relatively large and could even reverse the  sign of $R_H^{(0)}$. 
The lesson learned is that renormalization of $H$ onto the single band model  allows one to fully include the single body periodic potential  into the renormalized zeroth term of the Hall coefficient
and, therefore, {\em greatly reduce the magnitude of the correction term}.

{\em Comment on lifetime anisotropy:}   The Hall coefficient for the case of a $\bk$-dependent lifetime $\tau_\bk$ is given by Boltzmann equation~\cite{jones-zener,ziman,Ong} as
\bea
R_H^{\rm Boltz} &=&   2 {e^3   \over  c (\sigma^{\rm Boltz}_{xx})^2 } \int {d^d k \over (2\pi)^d} \left(-{\partial f \over \partial \epsilon}\right) F_\bk \tau_\bk^2 , \nonumber\\
 \sigma^{\rm Boltz}_{xx} &=& 2 e^2  \int {d^d k \over (2\pi)^d} \left(-{\partial f \over \partial \epsilon}\right)  |v^x_\bk  |^2 \tau_\bk .
\label{BS-tauk}
\eea
The  anisotropy  factor $(\tau^2_\bk-  \langle \tau\rangle^2 )/ \langle \tau\rangle^2 $ on the Fermi surface is a consequence of
anisotropic scattering by impurities, phonons, and other electrons. These effects are missing in $\bar{R}_H^{(0)}$ of Eq.~(\ref{BS1}), which depends only on the band structure.
Application of the fully interacting Liouvillian when constructing the higher order Krylov states introduces the anisotropies of the scattering operators, of the type shown in Eq.~(\ref{Corr-Boltz}).
However, at low temperatures and for weak scattering potentials,  Eq.~(\ref{Corr-Boltz}) is  simpler than computing $R_H^{\rm corr}$.  
Nevertheless, Eq.~(\ref{HallN}) teaches us that lifetime {\em anisotropy} effects can be described by equilibrium susceptibilities. 


The modified Nermst coefficient of a single band Hamiltonian in Eq. (\ref{H-SB}) is
\bea
W^{(0)} &=&    {2 e      \over  c \mu_0 \mu^Q_0 } \int {d^d k \over (2\pi)^d} \left(-{\partial f \over \partial \epsilon}\right)  (\epsilon_\bk-\mu)F_\bk ,\nonumber\\
\mu^Q_0  &=& 2 \int {d^d k \over (2\pi)^d} \left(-{\partial f \over \partial \epsilon}\right)  (\epsilon_\bk-\mu) |v^x_\bk  |^2 , \nonumber\\
&=&{ \pi^2\over 3}{k_B^2\over e^2 } T^2 \mu_0 +\cO(T^3) ,
\label{W-BS}
\eea
where, for the last line, we used a low temperature Sommerfeld expansion~\cite{behnia-book}.

Similarly, the  thermal Hall coefficient is given by
\bea
R_{TH}^{(0)} &=&  2 {e  T   \over  c   (\mu_Q)^2   } \int {d^d k \over (2\pi)^d} \left(-{\partial f \over \partial \epsilon}\right)  (\epsilon_\bk-\mu)^2 F_\bk. \nonumber\\
\label{W-BS}
\eea

Application of these results to the parabolic band  model from Eq. (\ref{parabolic}) yields the simple expressions
\bea
W^{(0)} &=& {1\over n \epsilon_F c},\nonumber\\
R_{TH}^{(0)}  &=& {3  e\over \pi^2 k_B  } ~{1\over n T c}.
\eea
For  parabolic bands, the inverse of $W$ ($R_{TH}$)  measures the number density times the Fermi energy (temperature).

 
\subsection{Hard Core Bosons (HCB)}
Repulsively interacting bosons in a deep periodic potential with square lattice symmetry are described by,
\be
H= \sum_i \left( {(\bp_i- {q\over c} \bA)^2 \over 2m} + V(\bx_i) \right)+ {1\over 2} \sum_{i\ne j} U(|\bx_i-\bx_j|).
\label{Hbosons}
\ee
This model may be renormalized onto a single-band, Bose-Hubbard model~\cite{BHM-Fisher}  (using $\hbar=c=1$)
\be
\bar{H} = -t \sum_{\langle ij\rangle}  e^{-i qA_{ij} }a^\dagger_i a_j + {\rm h.c.} + U \sum_i n_i^2 ,
\label{BHM}
\ee
where $a_i^\dagger$ creates a boson on site $i$, and $A_{ij} = \int_{\bx_i}^{\bx_j} d\bx\cdot \bA$.
At strong interactions, when the
average filling is between two integers $j <  \langle n_i\rangle   <j+1$,  the fluid phase is ``squeezed'' between two insulating phases.   
The effective Hamiltonian for that regime is well described by further renormalization onto the Hard Core Bosons  (HCB) model~\cite{lindner}
\be
\bar{H}^{\rm HCB} = -t \sum_{\langle ij\rangle } e^{-i q A_{ij} } S^+_i S^-_j + \mbox{h.c.} ,
\label{HCB}
\ee
where $\bf{S}$  are pseudospin half operators.  $S_i^+$ creates a HCB at site $i$, and $S_i^z\!=\!n_i-{1\over 2}$ measures its fluctuations in its occupation numbers.
The Hall coefficient vanishes at  $\langle n\rangle ={1\over 2}$ by emergent particle hole symmetry, which can be verified by  $S^+\to S^-$ and  $S^z\to -S^z$ in Eq. (\ref{HCB}).
The renormalized  currents  and magnetizations are
\bea
\bar{j}^\alpha &=&    - i  q t  \sum_i \left(  e^{-iq A_{ii+\alpha}} S^+_i S^-_{i+\alpha} -  \mbox{h.c.} \right) , \nonumber\\
\bar{M}&=&   {q\over 2   } \sum_{i }  x_i \bar{j}^y_{i+y} -  y_i \bar{j}^x_{i,i+x}.
\eea
Expanding Eq.(\ref{AB-susc}) in powers of $\beta$ at high temperatures yields
 \be
 (A|B) = \beta \Tr \rho_\infty A^\dagger B -  {\beta^2\over 2} \Tr \rho_\infty \{H,A^\dagger\} B + \cO(\beta^3) .
 \ee
The infinite temperature density matrix $\rho_\infty $ projects onto a fixed particle number  
\be
\sum_i \Tr \left( \rho_\infty  S^z_i \right) =(n-{1\over 2}) V .
\ee
Thus 
\be
\mu_0 \!= \!  \beta  \Tr \rho_\infty  j_{i,i+x}^2  . \ee
One can verify that all magnetization matrix elements $M_{2j,2k}''$ vanish unless the operators in the trace encircle a magnetic flux. 
Therefore, for a {\em triangular} lattice at high temperatures~\cite{SSS},   $M_{0,0}''  \!\propto \! - \beta   (n-{1\over 2})$, while
for a square lattice,  $M''_{0,0}\!\propto\! -\beta^2 (n-{1\over 2})$. 
Thus we obtain for the triangular and square lattices 
\be
\bar{R}_H^{(0)} \propto  \left\{ \begin{array}{ll}- T (n-{1\over 2}) & \mbox{triangular}\\  -  (n-{1\over 2}) & \mbox{square}\\ \end{array} \right.
~.
\label{Sign}
\ee
Correction terms that involve $\bar{M}_{2j,2k}''$  
 decay  rapidly with $j,k$ due to diminishing overlaps between Krylov states. Thus the Hall sign changes around half filling lines are denoted by HCB in Fig.~\ref{Fig:LB}.\\

 \subsection{Quantum Rotors (QR)}
For the same Bose-Hubbard model in Eq. (\ref{BHM}) near  the Mott phases at integer filling $n_0$, the   
fluid state can be described by the Quantum Rotors (QR) field theory 
\begin{widetext}
 \be
\bar{H}^{\rm QR} = \int d^d x  {1\over 2\chi_c} (\rho(\bx)-n_0 a^{-d})^2+ {1\over 2} \rho_s \left(\nabla\varphi(\bx) +{q\over c} \bA\right)^2 (1+ \gamma \rho(\bx)^2) +  V(\bx)\rho(\bx) ,
\label{Hrotor}
\ee
\end{widetext}
where $a$ is the lattice constant, $\chi_c$ is the local compressibility, and $\rho_s$ is the local superfluid stiffness.  $\rho$ is the deviation of the density from the commensurate filling $n_0$ of the neighboring Mott phase. The QR theory can be derived from a quantum Josephson junction array model, where $\chi_c$ are the grain capacitances, and  $\rho_s$ 
are intergrain Josephson couplings.  

From the phase diagram of the Bose-Hubbard model, it is clear that $\gamma >0 $, since the superfluid stiffness and ground state order parameter are enhanced
as the density is varied away from  $n_0$.

The canonical density-phase commutations are~\cite{Com3} 
\be
[\rho(\bx),\varphi(\bx')] = -i \delta(\bx-\bx') .
\label{rhophi}
\ee
The  QR currents and magnetization densities are
\bea
\bar{\bj} (\bx) &=& - q \rho_s  \nabla\varphi  (1+ \gamma \rho^2) ,\nonumber\\
\bar{m}(\bx)&=& - {q\over 2c} \left(  x j^y(\bx) -   y j^x(\bx)\right) .
\eea
Notice that the factors $\gamma \rho^2$ are necessary to produce current dynamics via  nonvanishing  commutators $\cM j^\alpha$ and  $\cL j^\alpha$. There is no Hall effect 
at the particle-hole symmetric line $\langle \rho\rangle =0$.

Using Eq. (\ref{rhophi}),  the sign of the Hall coefficient  can be obtained
\bea
R_H^{(0)}  \propto      {\gamma  \over q \rho_s c }  \langle  \rho   \rangle + \cO (\langle \rho^3\rangle) ,
 \label{QR}
 \eea
 which implies a particle-like Hall effect above the commensurate filling, and a hole-like effect below.
As for the band electrons, higher order corrections of $R_H^{\rm corr}$  are negligible at weak disorder.

\begin{figure}[!t]
\begin{center}
\includegraphics[width=8.5cm,angle=0]{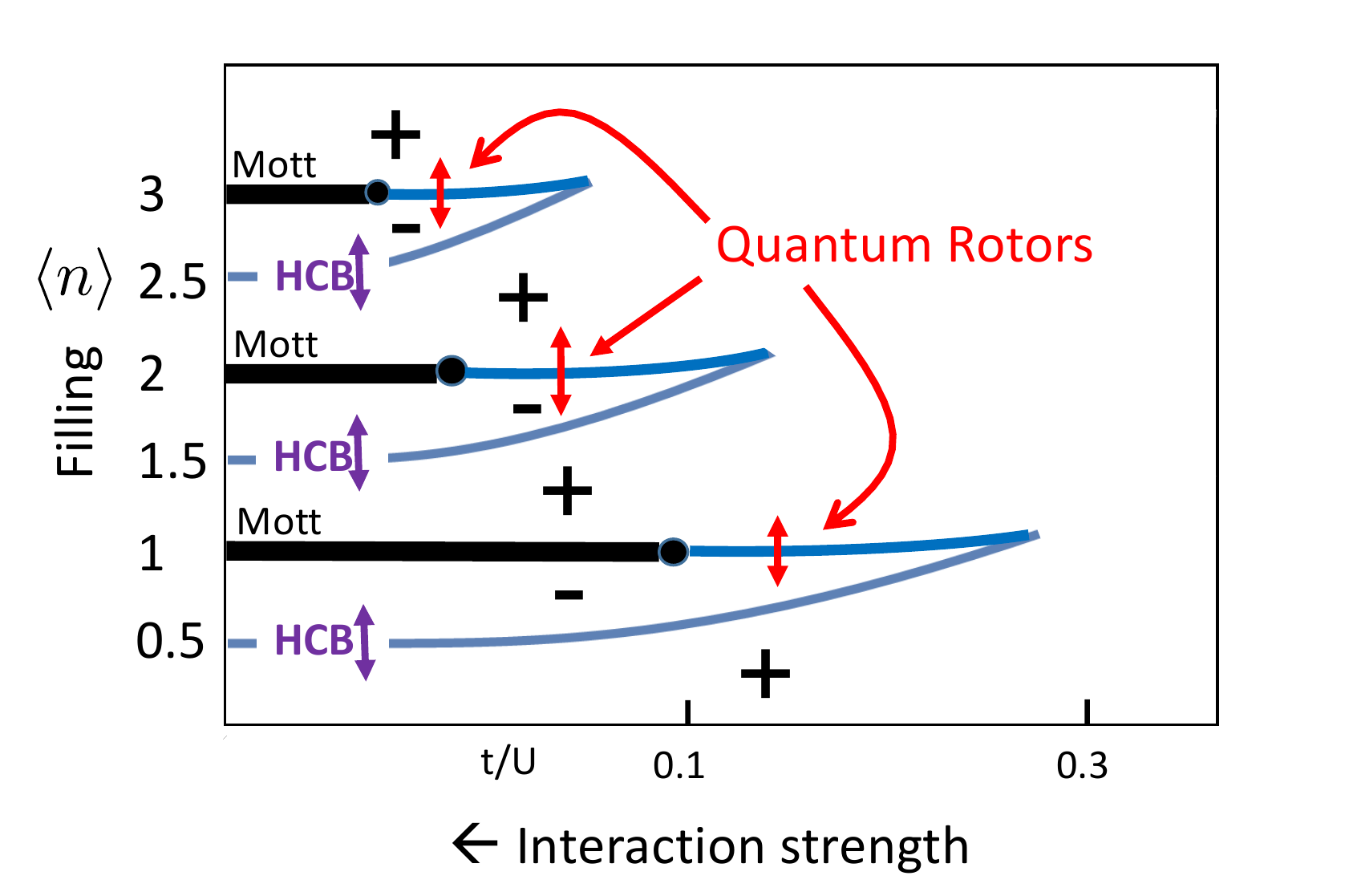}
\caption{Hall signs in strong interactions regime of the Bose-Hubbard model Eq.~(\ref{BHM}).  Mott insulators are thick black lines, ending at critical points (black circles). 
Solid blue lines mark Hall sign changes at zero temperature, computed by Huber and Lindner\cite{huber}.
At high temperatures, we find the same sign changes using Hard Core Bosons (HCB) and Quantum Rotors,  in Eqs.~(\ref{Sign}) and (\ref{QR},) respectively.
}
\label{Fig:LB}
\end{center}
\end{figure}

In Fig.~\ref{Fig:LB}  we combine the results of HCB and QR models to map the Hall signs of the Bose-Hubbard model in the nonsuperfluid (metallic) phase.
While the Hall conductivity of metallic phases are not simply related to Chern numbers on finite tori (see Section \ref{Sec:Other}),
it is interesting that our results in Fig.~\ref{Fig:LB} are consistent with the Hall signs as evaluated by Huber and Lindner~\cite{huber}.   

\section{Streda formulas, Chern Numbers and Hall-pumped Polarization }
 \label{Sec:Other}
This section is peripheral to the bulk of the paper and is included for completeness of our discussion of equilibrium magneto-transport coefficients.
 We derive some previously  known relations for Hall and TTE conductivities, which are  applicable to nondissipative phases and high magnetic fields.

\subsection{Translationally invariant systems}
 
The Hall coefficient of a  perfectly translationally invariant system~\cite{Girvin,Cooper} subject to a uniform electric field $\bE=E^y \hat{\by}$ is readily solved
by a  Galilean transformation to a moving frame of velocity
\be
v^x = c {E_y \over B } ,
\ee
where the electric field is transformed to zero in the moving frame. Absence of moving potentials implies that the current vanishes in the moving frame.
Hence, back  in the lab frame the current is
\be
\langle j^x \rangle = qn\bv =\sigma_H E^y~~\leadsto~ \sigma^{\rm hom}_H = { nqc\over B} ,
\label{sxy-GI}
\ee 
where $nq$ is the charge density, and
\be
\langle j^x_Q \rangle = T {S\over N} n v^x  =T \alpha_{xy} E^y~~\leadsto~\alpha^{\rm hom}_{xy} = { s \over B} ,
\label{axy-GI}
\ee
where $s= S/\cV$ is the entropy density~\cite{behnia-review}.
Eqs.~(\ref{sxy-GI}) and (\ref{axy-GI}) apply  at any density, magnetic field and two-body interactions, as long as there are no spatially varying potentials, 
and consequenty zero resistivity.

\subsection{Streda formulas for $\sigma_H$ and $\alpha_{xy}$}
An equilibrium formula for the Hall conductivity was proposed by Streda~\cite{Streda}
\be
\tilde{\sigma}_{H} = c   \left({\partial \rho \over \partial B}\right)_{\mu,T} =c   \left({\partial m\over \partial \mu }\right)_{\rho,T},
\label{Streda1}
\ee
where $\rho$ and $m$ are the charge and magnetization density respectively.

For the TTE, a similar Streda formula  is
\be
\tilde{\alpha}_{xy}  =  c  \left( {\partial s \over \partial B}\right)_{\mu,T}.
\label{Thermal-Streda1}
 \ee
 
Here we show that both Streda formulas are related to the {\em static long wavelength conductivities}. Note that these imply the reverse order of limits than
the DC limit in  Eq~(\ref{limits}). That is to say, 
\bea
\tilde{\sigma}_{H}  &=&  \lim_{\bq\to 0}  \sigma_{H} (\bq,0)\nonumber\\
 \tilde{\alpha}_{xy} &=& \lim_{\bq\to 0}   \alpha_{xy} (\bq, 0).
\label{Thermal-Streda2}
\eea

{\bf Proof}:  The continuity equation relates charge density $ \rho $ to current density  
 \be
  \dot{\rho}(\bx) = {i\over \hbar} \cL \rho  =  -\nabla \cdot \bj (\bx) .
 \ee
By a  Fourier transformation 
 \be
 {1 \over \hbar} \cL \rho_\bq =   -   \bq\cdot \bj_\bq ,
\label{cont}
\ee
the  relation between magnetization (in the $z$ direction) and magnetization currents $\bj_m$  is
\be
\nabla \times {\bf M} = {1\over c} \bj_{m} .
\ee
Without loss of generality, we choose  $\bq= (q_x,0)$, $\bj_{m}= (0,j^y )$, and  ${\bf M}_\bq  = M_\bq \hat{\bz}$. Thus we can write 
\bea
j^x_\bq  &=& -{1\over  \hbar q_x } \cL \rho_\bq    \nonumber\\
\to \left({1\over \cL}\right)'  j_\bq^x&=& - {1\over \hbar q_x} \rho_\bq ,
\label{jq-nq}
\eea
and
\be
j^y_\bq   =  i  c q_x M_\bq  .
\label{jq-Mq} 
\ee
Using Eqs. (\ref{Kubo}), (\ref{jq-nq}), (\ref{jq-Mq}), we obtain
\bea
\lim_{\bq\to 0}   \sigma_{H}(\bq,0) &=&    \lim_{\bq\to 0} \lim_{\cV\to \infty} {1\over   \cV }\Re \Big( \rho_\bq \Big| M_{\bq}\Big) ,\nonumber\\
&=& c   \left( {\partial \rho \over \partial B}\right)_{\mu,T} .
 \eea

Similarly, using a Fourier transform of Eq.~(\ref{jQ-Qx}) for the  TTE coefficient yields
\be
j_Q^x(\bq) ={1\over \hbar q_x }  \cL ( h_\bq -\mu n_\bq ) .
\label{jQ-hq}
\ee
Rewriting Eq. (\ref{Kubo-alphaxy}) using Eqs. (\ref{jq-Mq}) and (\ref{jQ-hq}) yields
\be
\lim_{\omega\to 0} \alpha_{xy}(\bq,\omega) =   {c   \over  T \cV    }\left(  \left( h_\bq-\mu n_\bq \Big| M_\bq\right) -   \langle M_\bq   \rangle\right) .
 \label{alphaxy-q}
\ee
Taking the limit $\bq \to 0$ of Eq.~(\ref{alphaxy-q}) and using the equilibrium relation
\be
\left( {d (E-\mu N - TS   ) \over dB} \right)_{\mu,T}= \langle M \rangle ,
\ee
where $S$ is the entropy, we obtain
\bea
\tilde{\alpha}_{xy}\equiv \lim_{\bq\to 0}   \lim_{\cV\to \infty}  \alpha_{xy}(\bq,0) =     c  \left( {\partial s \over \partial B}\right)_{\mu,T} ,
\label{Thermal-Streda}
 \eea
 where $s= {S\over \cV} $ is the entropy density, which completes the proof of Eq.~(\ref{Thermal-Streda2}). Q.E.D.

Eq.~(\ref{Thermal-Streda2}) allows us to investigate sufficient conditions for permitting reversal of order-of-limits, required by Eq.~(\ref{limits}).  
If there exists an equilibrium  gap  $E_{\rm gap}  = \lim_{\bq\!=\!0}{\rm min}_n(E_n(\bq)-E_0)\gg T >0$ 
which survives the  limit of $\cV\to \infty$, surely the order of limits can be reversed. This is permitted in  quantum Hall phases, where the only gapless regions are at the sample edges~\cite{Comm-Streda}.  
On the other hand, metals at weak magnetic fields are gapless in the bulk, and not described by the Streda formulas.
 
\subsection{Chern numbers on the torus}
A finite gauged torus is penetrated by a uniform magnetic field with integer total flux $N_\Phi \Phi_0$.  Here, $\Phi_0=hc/q$, where $q$ is the charge of the particles.
 Its two holes are threaded by Aharonov-Bohm  fluxes ${\theta_\alpha \over 2\pi}\Phi_0, \alpha=x,y$. 

Aharonov-Bohm (AB) fluxes can be introduced by adding source terms to the Hamiltonian   
 \be
 H\to H -   {\hbar \over q}  \left( j^x \theta_x /L_x +    j^y  \theta_y/L_y \right) .
 \label{Htheta}
 \ee
On the torus, we cannot define polarization operators. Nevertheless, we can relate write the matrix elements of $\left({1\over \cL}\right)' j^\alpha$ 
using  first-order perturbation theory in $\theta_y$ to eigenstate $|n\rangle$, as
  \bea
\langle m|  \left( {1\over \cL}\right)' j^\alpha |n\rangle &=&     { \langle m| j^x  |n\rangle \over  E_m-E_n} , \nonumber\\
&=&   { L_x q\over \hbar}   \Big\langle m| {\partial\over \partial \theta_x} n \Big\rangle_{\vec{\theta}=0}  .
\label{jmn}
 \eea
 Substituting Eq. (\ref{jmn}) into Eq. (\ref{Kubo-sxy-general}), the Hall {\em conductance} of the torus is
 \be
\Sigma_{H} (L_x,L_y) =   2 {q^2 \over \hbar   }   
\sum_{n=0}^\infty  \rho_n  \Im  \Big\langle  {\partial\over \partial \theta_y} \psi_n   \Big|  {\partial\over \partial \theta_x} \psi_n \Big\rangle_{\vec{\theta}=0}  ,
\label{ChernCurve}
 \ee
which is the thermally averaged {\em Chern curvature} at zero AB fluxes. 

Avron and Seiler \cite{yosi}, using adiabatic transport theory,  related the ground state Hall conductance to the integral of the Chern curvature over the AB fluxes (the reciprocal torus):
 \bea
 \Sigma^{\rm Chern}_{H}   &=&   \int_0^{2\pi}  \int_0^{2\pi} {d\theta_x d\theta_y \over (2\pi)^2 }~ \Sigma_{H}(\theta_x,\theta_y, T\!=\!0 ) , \nonumber\\
 &=& {q^2\over h}  \int_0^{2\pi}  \int_0^{2\pi}  {d\theta_x d\theta_y \over \pi  }~ \Im  \Big\langle  {\partial\over \partial \theta_y} \psi_0   \Big|  {\partial\over \partial \theta_x} \psi_0\Big\rangle , \nonumber\\
 &=& {q^2\over h} \times\mbox{\rm Integer} .
 \label{Chern}
\eea

The  double integral over a smooth Chern curvature yields a topological integer called Chern  number~\cite{TKNN,yosi}. In the limit of large tori with a finite gap, the
Chern curvature at weak magnetic field is expected to approach its average, and the two expressions in Eqs. (\ref{ChernCurve}) and (\ref{Chern}) coincide.

The important conclusion from relating the Chern number to $\Sigma_H$ is that the Hall conductance is quantized as long as the conditions of adiabatic transport theory hold.  
Eq.~(\ref{ChernCurve}) is a static equilibrium calculation. At low temperatures, it  requires only the knowledge of  the lowest eigenstates~\cite{lindnerHall}.

Huber and Lindner (HL)~\cite{huber} proved an important theorem about ground state Chern numbers of charged particles in periodic potentials.\newline
{\bf HL Theorem:  } Consider $N$  particles (fermions or bosons) of charge $q$ on the surface of a torus,
in a periodic potential of $N_{\rm sites}$ unit cells, and a perpendicular uniform magnetic field of comensurate flux $N_\phi \Phi_0$, where $ N_{\rm sites}/N_\phi$ is integer. 
\be
\Sigma^{\rm Chern}_{H}  = {q^2\over h} \left( \nu +  m { N_{\rm sites} \over N_\phi} \right), 
\label{HL}
\ee
where $\nu = {N \over N_\phi}$ is the filling factor, and $m$ is any integer.  \\

{\bf Proof:} Define a flux quantum cell of size $(L^{\Phi_0}_x,L^{\Phi_0}_y )$, such that $L^{\Phi_0}_x \times L^{\Phi_0}_y = N_{\rm sites}/N_\phi$.
Because of the relation between translations of the null lines of the vector potential in $H$~\cite{lindnerHall} and changes in the AB fluxes, 
the Chern curvatures (which are gauge invariant) are periodic in the AB fluxes with 
the corresponding periodicity 
$\Delta\theta_x =  2\pi / L^{\Phi_0}_x$, $\Delta\theta_y =  2\pi /  L^{\Phi_0}_y$ respectively.
Any change in the Chern number can occur by a level crossing, which can introduce an integer change in the  total Chern number   $m=\pm 1,\pm2 \ldots$. 

We first consider a free Hamiltonian with zero potential energy.  Galilean symmetry requires
\be
\Sigma^{\rm Free}_{H}  = {nqc\over B} = {q^2 \over h} \nu .
\ee
By the argument above, turning on the periodic potential adiabatically can only change  the Chern number by an integer $m$ multiplied by the number of periodic flux quanta unit cells $N_{\rm }/N_\phi $, which results in Eq.~(\ref{HL}). QED.

A change with  $m= -1$  reverses the Hall sign, which is expected above half filling 
for  HCB \cite{lindnerHall}.  For noninteracting tight binding electrons on a bipartite lattice~\cite{berg}, $m=-2$ across the half filling boundary.

\begin{figure}[!t]
\begin{center}
\includegraphics[width=8.5cm,angle=0]{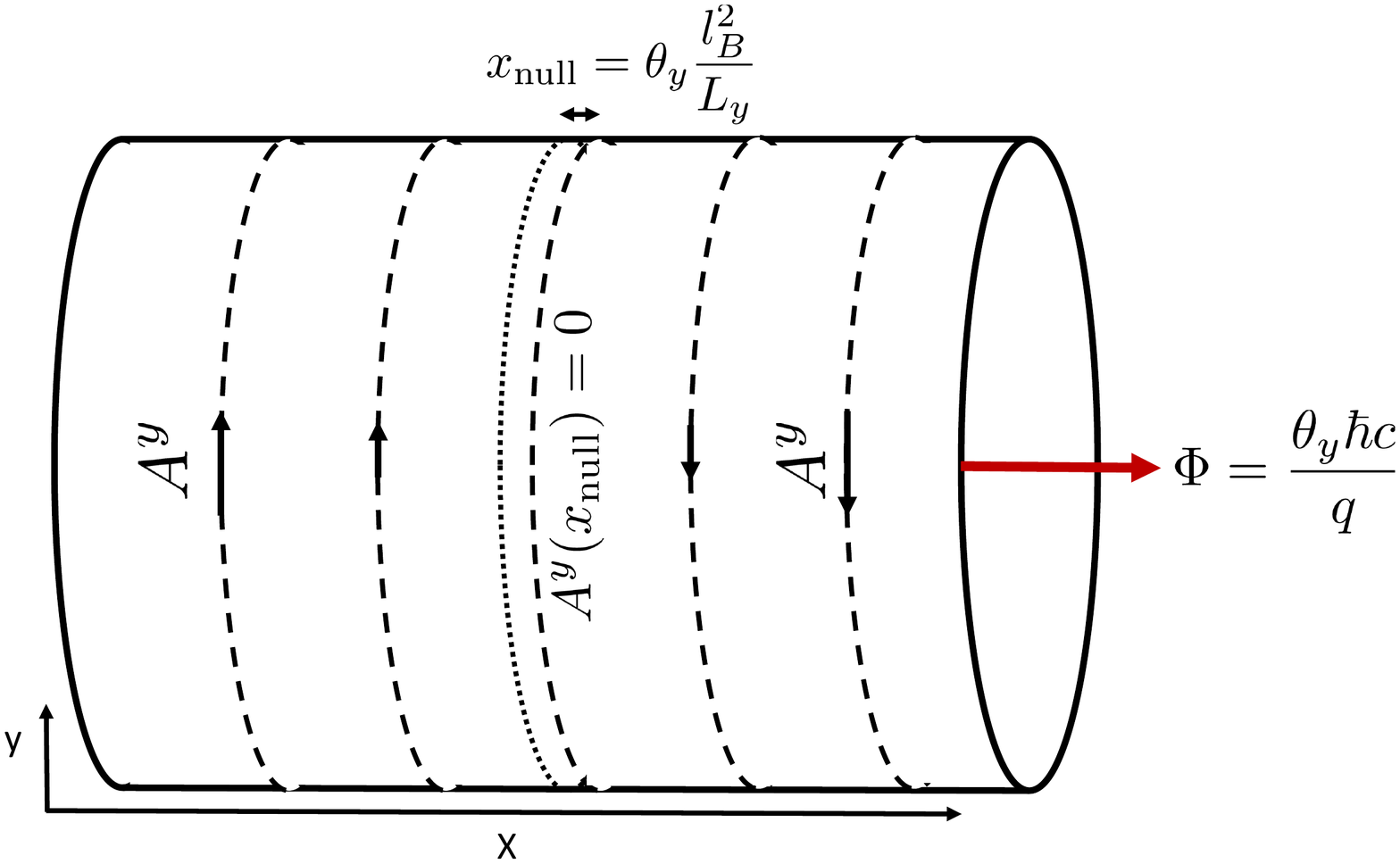}
\caption{The gauged cylinder. A finite cylinder penetrated by a radial magnetic field, and Aharonov-Bohm flux $\Phi$. $L_y$ is the circumference, and $l_B = \sqrt{\hbar c\over qB}$ is the magnetic length. $A^y$ is the vector potential, 
whose null line at $x_{\rm null}$ moves as a function of $\theta_y$, whilst pumping the charge polarization.
}
\label{Fig:cylinder}
\end{center}
\end{figure}

\subsection{Hall-pumped polarization on the cylinder}
Hall conductance on a finite cylinder is related to the  Hall-pumped polarization~\cite{Chern-pump-theory,prelov-ladder}.
We assume periodicity of $H$ in the $y$-direction and open boundary conditions on the $x$ axis (see Fig.~\ref{Fig:cylinder}).  For charge $q$ particles, $x$-polarization is
\be
P^x=  q\sum_i x_i .
\ee
A small AB flux $\theta_y \hbar c/q$ is introduced through the cylinder's hole by adding to the Hamiltonian
\be
 H\to H -  {\hbar \over q L_y }   j^y  \theta_y .
 \label{Htheta}
 \ee
Inserting Eqs. (\ref{Kubo-sxy-general}) and (\ref{P-tP}) into Eq. (\ref{jmn}), the cylinder's Hall conductance is given by
\bea
\Sigma^{\rm pump}_H     &=& { q \over \hbar L_x } \sum_{n=0}^\infty  \rho_n   \nonumber\\
&&~\times  \left( \langle  \psi_n   | P^x | {d\over d\theta_y } \psi_n \rangle  +  \langle  {d\over d\theta_y } \psi_n   | P^x |  \psi_n \rangle  \right) , \nonumber\\
 &=& { q   \over  \hbar  L_x  }  \sum_{n=0}^\infty  \rho_n    { d P^x (n,\theta_y) \over d \theta_y} \Big|_{\theta_y=0} , \nonumber\\
 & = & { q   \over  \hbar  L_x  } {  d\over d\theta_y}\langle P^x\rangle\Big|_{\theta_y=0} .
 \label{pump}
 \eea
 This result can be interpreted as adiabatic pumping of the polarization, as depicted in Fig.~\ref{Fig:cylinder}.
 By reflection symmetry, we can define  $\langle P^x  \rangle=0$ for $\theta_y=0$.  The  {\em ``null line''}  at $x=x_{\rm null}$~\cite{lindnerHall} is defined by vanishing Wilson loop  $\oint dy A_y(x_{\rm null})=0$.
Adiabatically increasing $\theta_y\to  \Delta\theta_y$ moves the null line  an incremental distance 
\be
\Delta x_{\rm null} = \Delta \theta_y { L^x \over 2\pi N_\phi}
\ee
along the $x$ axis. If there are non level crossings, 
the variation of Hamiltonian and its eigenstates adiabatically pumps the  polarization. If the pumping takes time $\tau$, the $x$-current is given by $I_x =  {1\over L_x}  \Delta  \langle P^x \rangle /\tau$, and
the $y$-voltage is $V^y = {\hbar  \Delta \theta_y \over q\tau}$, which yields Eq.~(\ref{pump}) for $\Sigma^{\rm pump}_H=  I_x/V_y$. 

$P^x(\theta_y)$  is a thermodynamic average which can be  computed at any fixed $\theta_y$ by equilibrium approaches. For example, using
a variational matrix product state as provided by  e.g.  Density Matrix Renormalization group~\cite{DMRG}.

{\em Caveat:} The relevance of Chern curvatures and numbers  and Hall-pumped polarization to the limit $\cV\to \infty$ 
depends on an absence of level crossings for infinitesimal changes of AB fluxes $\Delta\theta$.  These can give rise to dissipative relaxation of the polarization.
The incompressible quantum Hall phases satisfy this condition. Their polarization can only relax
by charge tunneling between far away edge excitations whose rate is  suppressed exponentially in the distance between edges, for 
both integer and fractional quantum Hall phases~\cite{Assa-QH}. 
However,  adaibatic transport fails for bulk-gapless disordered metals, where
{\em nonadiabatic}  (Zener tunneling) at arbitrary weak electric field gives rise to a dissipative conductivity $\sigma_{xx}  >0$.

\section{Summary and Discussion}

The main purpose of this paper was to derive  formulas for the Hall, modified Nernst, and thermal Hall coefficients, which avoid computing DC conductivities.
Quite remarkably, these coefficients for dissipative metals of $\sigma_{xx}>0$  depend on the free energy and its {\em static} derivatives. 
As such, they are now amenable to a variety of well-developed numerical methods, which could be applied to interesting
models of strongly interacting electrons and bosons, such as the Hubbard and t-J models~\cite{IEQM,Ilia} for cuprates and metals near Mott insulators,  the Kondo latice model for heavy fermions~\cite{hewson},  
Weyl semimetals~\cite{ari-weyl}, cold  atoms on optical lattices with an artificial magnetic field~\cite{Chern-pump-exp}, and more.

The zeroth terms $R_H^{(0)}$, $W^{(0)}$, and $R_{TH}^{(0)}$ are relatively simple and can be evaluated analytically in certain models and limits (e.g. weak interactions,
or lattice models at high temperature). The correction terms 
require  susceptibilities of  more complicated operators. Since the sums are expected to converge for noncritical metals, the higher order terms
should decrease in magnitude, but the rate depends on the model and temperature regime. 
In practice, the first few  terms could provide an estimate of the convergence rate and the truncation error.

As our examples show in Section \ref{Sec:Appl},  large potential variations and two body interactions may be renormalized 
at low energies into simpler effective Hamiltonians. The single band model for weakly interacting electrons
and hard core bosons and quantum rotors for strongly interacting  bosons are such examples. By renormalization,  {\em qualitative}  features of magneto-transport coefficients,  such as sign changes,  temperature, and doping dependences, may be extracted 
already from  the zeroth order coefficients.

{\em Strong disorder:}    $R_H$ in disordered metals near the localization transition have been extensively studied. 
For noninteracting electrons in two dimensions, microscopic calculations~\cite{Fuku,Altshuler} have shown that $R_H$ 
remains constant,  while the longitudinal and Hall conductivities vanish at low temperature.
In three dimensions,  scaling arguments near the mobility gap~\cite{Boris}, have also shown that $\sigma_{xx}^2 \sim \sigma_H$ vanish, while $R_H$ is remains constant at the metal to insulator transition. The Hall resistivity of  the Puddle Network Model (a network of quantum Hall puddles of a fixed filling factor, connected by arbitrary resistors) was shown to be independent of the longitudinal  resistivity~\cite{AS}. These results could  be  interpreted as the insensitivity of $R_H$ to relaxation rates and wavefunction localization. 
In two dimensions, effects of interactions have been found to give rise to logarithmic divergence of $R_H$ at low temperatures~\cite{Altshuler}. 
It would be interesting to investigate within our formula 
which equilibrium susceptibilities  are responsible for the diverging Hall coefficient at low temperatures.

{\em Acknowledgements.}  I thank Yosi Avron, Noga Bashan,  Kamran Behnia, Snir Gazit, Duncan Haldane, Bert Halperin, Ilia Khait,  Ganpathy Murthy, Boris Shapiro, Efrat Shimshoni and Ari Turner, for useful discussions. 
I acknowledge support from the
US-Israel Binational Science Foundation grant 2016168 and  the Israel Science Foundation grant 2021367. I thank the Aspen Center for Physics,  grant NSF-PHY-1066293, and Kavli Institute for Theoretical Physics at Santa Barbara, where parts of this work were done.

\appendix
 
\section{Krylov states}
\label{App:Krylov}\begin{figure}[!b]
\begin{center}
\includegraphics[width=7.5cm,angle=0]{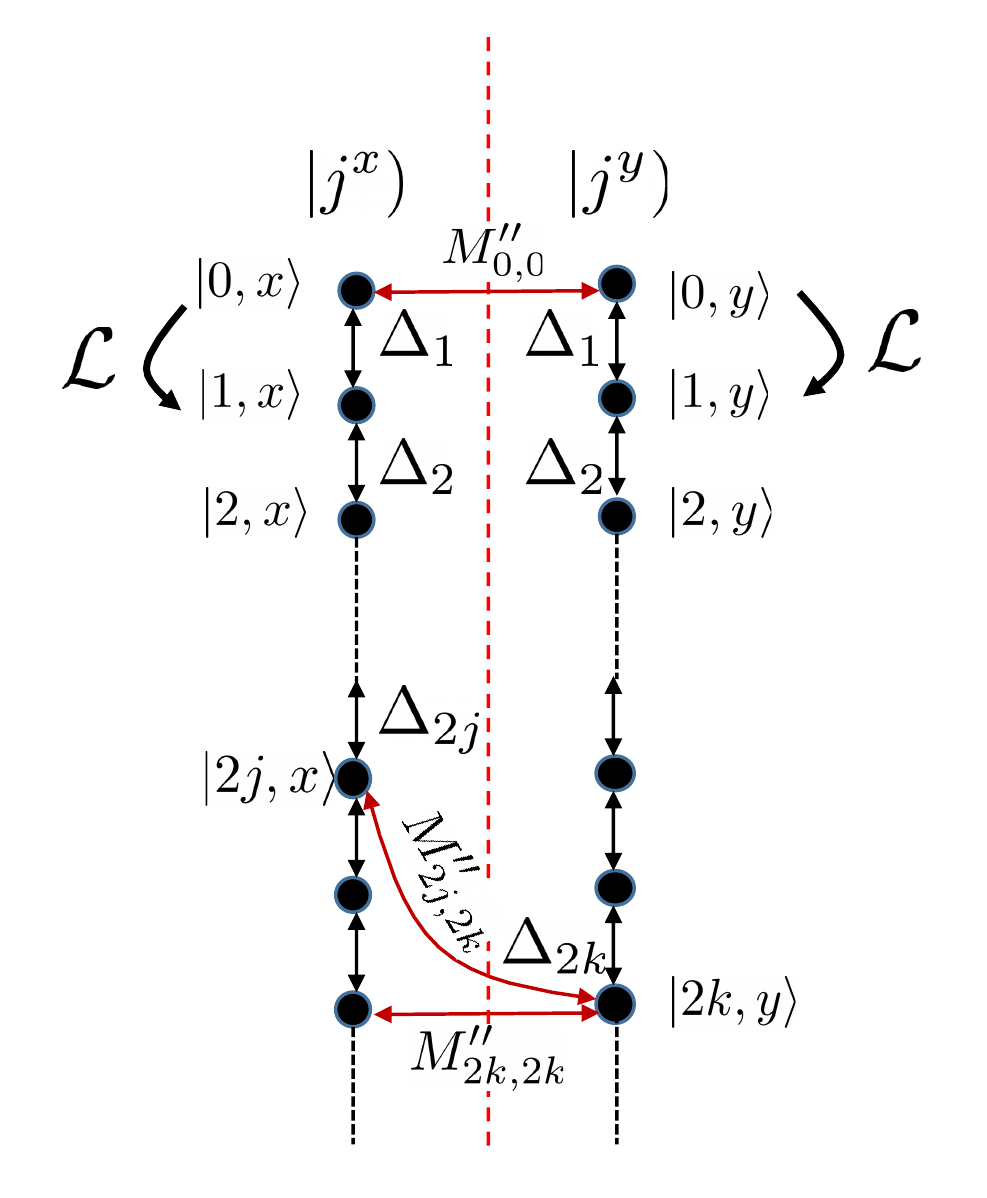}
\caption{The orthonormal Krylov bases, Eq.~(\ref{Krylov}), constructed for $B\!=\!0$ from $j^x$ and $j^y$ by repeated application of the Liouvillian $\cL$. 
$\Delta_n$ are the recurrents of $\sigma_{xx}$. 
$M''_{n,m}= \Im \langle n/_{j^x}| \cM | m/_{j^y}\rangle$ are the magnetization matrix elements used  in Eq.~(\ref{dsdB}).
}
\label{fig:Krylov}
\end{center}
\end{figure}

Bogoliubov hyperspace is the Hilbert space of operators (hyperstates) $|A),|B)$. The inner product given by Eq.~(\ref{product1}) depends on the Hamiltonian $H$ and inverse temperature $\beta$ and can be written in several forms
\bea
(A|B) &=&  \sum_{nm} {\rho_n-\rho_m\over E_m-E_n} \langle m |A^\dagger|n\rangle \langle n | B|m\rangle , \nonumber\\
 &=& -\partial_{h_A} \partial_{h_B} \Tr \log e^{- (\beta H -h_A A -h_B B)} \Big|_{h_A,h_B\!=\!0} , \nonumber\\
 &=& \int_0^\beta d\tau \langle A(\tau) B\rangle .
\label{product}
\eea
The first line can be used to confirm that $(A|B)=(B|A)^*$ and $(A|A)\ge 0$. 
The second line shows that  $(A|B)$ is  an equilibrium  susceptibility obtained by adding static source terms $-h_A A -h_B B$   to $H$ before differentiation.
 The third line relates the inner product to an imaginary time correlation function. 
Here, $\langle O \rangle \equiv \Tr \rho O$, and $A(\tau) \equiv e^{H\tau} A e^{-H\tau}$.  Where possible,  $(A|B)$ could be computed by imaginary time quantum Monte Carlo algorithms~\cite{prokofiev,assaad}.

The Liouvillian is a hermitain hyperoperator
\be  
 \cL |A) = |[H,A]) .
\ee
The hyperresolvent requires an "$i\ve$'' prescription, which defines the hermitian and antihermitian parts as
\bea
&&\left( { 1\over \cL-\omega -i\ve}\right)  \equiv   \left( { 1\over \cL-\omega  }\right)' +i \left( { 1\over \cL-\omega  }\right)''\nonumber\\
&& \equiv   \left( { \cL-\omega \over ( \cL-\omega)^2 + \ve^2   }\right) + i\left( { \ve \over (\cL-\omega)^2 +\ve^2  }\right).
\eea
According to Eq. (\ref{limits}) we must keep $\ve>0$ as we take $V\to \infty$.

We  construct an orthonormal Krylov basis of operators, which will allow a  matrix representation of the Liouvillian and its inverse.
We start with the normalized root state
\be
|0/_A\rangle \equiv  {|A) \over \sqrt(A|A) } ,
\ee
where $A$ is the root operator (i.e. the uniform electrical or thermal current, in this paper).

Assuming $A$ is not in the kernel of $\cL$, we  construct the Krylov basis as follows
\bea
|1/_A \rangle &=& \cL|0/_A\rangle  ,  \nonumber\\
|n/_A)  &\equiv&    (1-\cP_{n-2}) \cL  (1-\cP_{n-3}) \cL \cdots (1-\cP_0) \cL^2 |0/_A\rangle, \nonumber\\
|n/_A\rangle &=&N_n |n/_A ),\nonumber\\
N_n &=&{1\over \sqrt{ (n/_A|n/_A) } },
\label{Krylov}
\eea
where $\cP_n = |n/_A\rangle \langle n/_A|$.

It is easy to verify that the Krylov basis is  orthonormal
\be
\langle n/_A|m/_A\rangle = \delta_{mn}
\ee
and can be used to span the subspace  $\cS_A= \{ \cL^{n}|A) \}_{n=0}^\infty$ by the resolution of identity in the subspace $\cS_A$
\be
\sum_{n=0}^\infty |n/_A\rangle\langle n/_A| = 1_{\cS_A} .
\label{ROI}
\ee

Henceforth, we drop the label ``$/_A$'' in the hyperstates, unless needed. 

The matrix representation of the Liouvillian  in this  Krylov basis is
\be
  \langle n |\cL| m \rangle \equiv L_{nm}   =    \left( \begin{array}{cccc}
  0&     \Delta_1& 0 & \ldots\\
   \Delta_1& 0  &    \Delta_2&  \ldots\\
0&    \Delta_2&  0 &   \\
\vdots& \vdots &   &\ddots
\end{array}
\right)_{nm} ,
\label{Lmn}
\ee
where  $\Delta_n, n=1,2,\ldots$ are the {\em recurrents}, which are calculated in Appendix \ref{App:mom-rec}.

If both $A$ and $B$ are either hermitian or antihermitian, $(A|B)$ is purely real. 
If we choose  $A$ to be hermitian, $\cL^{2j}A$ ($\cL^{2j+1} A$) is hermitian (antihermitian) for $j=0,1,\ldots$.  Hence, 
$|2j \rangle$ ($|2j+1\rangle$) are hermitian (antihermitian), and 
$\Delta_n=\langle n+1| \cL | n\rangle$ are purely real.

The Liouvillian Green function $\langle n| \left({1\over \cL -z}\right) |m\rangle$ is the inverse of a tridiagonal matrix
\bea
G_{n, m}(z) &=&- \langle n| \left({1\over z-\cL}\right)|m\rangle , \nonumber\\
&=&   \left( \begin{array}{cccc}
  -z&    \Delta_1& 0 & \ldots\\
  \Delta_1& -z  &   \Delta_2&  \ldots\\
0&   \Delta_2&  -z &   \\
\vdots& \vdots &   &\ddots
\end{array}
\right)^{-1}_{n,m} .
\label{Gz}
\eea

\section{Continued Fraction of Longitudinal Conductivities}
\label{App:CF}
The $(0,0)$ value of Eq.~(\ref{Gz}) is an infinite continued fraction
\be
G_{0,0} (z) =  - { 1 \over  z -  {\Delta_1^2 \over  z- {\Delta_2^2 \over z - {\Delta_3^2\over\vdots}}}} .
\label{Gz1}
\ee

The dynamical longitudinal dynamical conductivities are given by
\bea
\sigma_{\alpha\alpha}(\omega)  &\equiv&    \hbar \mu^\alpha_0~G''(\omega)_{0,0} , \nonumber\\
&=&  -\hbar \omega \Im {1 \over \hbar\omega+ i\ve -  {|\Delta_1|^2 \over \hbar\omega+ i\ve- {|\Delta_2|^2 \over  \hbar\omega+ i\ve -{\over \vdots}}}} .
\label{CF-sigma}
\eea
where 
\bea
\mu_0 &=&  {1\over \cV} (j^x| j^x  ) = \int_{-\infty}^\infty \! { d\omega\over 2\pi}\sigma_{xx}(\omega) ,\nonumber\\
\mu^Q_0 &=&  {1\over  \cV} (j_Q^x | j^x_Q  )  =  \int_{-\infty}^\infty \! { d\omega\over 2\pi}\kappa_{xx}(\omega)
\eea
are the zeroth moments (sum rules) of the conductivity and the thermal conductivity respectively.

For   continuum particles of charge $q$, mass $m$, and density $n$  
\be
\mu_0 = {1\over \hbar V} \Tr \rho [P^x,j^x] = {nq^2\over m} ,
\ee
which is known as the f-sum rule.
The thermal conductivity sum rule is given by Eq. (\ref{trace}) as
\be
\mu_0^Q = {1\over \hbar V} \Tr \rho [Q^x, j_Q^x]  .
\ee
This sum rule was  introduced as $\Theta_{xx}/T$ and evaluated by Shastry~\cite{shastry-sumrules} for certain models.

The DC order of limit (\ref{limits}) of (\ref{CF-sigma}) is
\bea
\sigma_{xx}  &=&  \Im {-\hbar \mu_0 \over  i\ve -  {|\Delta_1|^2 \over   i\ve- {|\Delta_2|^2 \over    i\ve -{\over \vdots}}}},\nonumber\\
\kappa_{xx}&=&  {1\over T}\Im {-\hbar \mu^Q_0 \over  i\ve -  {|\Delta^Q_1|^2 \over   i\ve- {|\Delta^Q_2|^2 \over    i\ve -{\over \vdots}}}} .
\label{CF-sigma}
\eea
 
\section{Computing recurrents from moments}
\label{App:mom-rec}
Here we show how the recurrents $\Delta_n, \Delta_n^Q,~ n=1,2,\ldots$ can be computed recursively from their respective moments, which 
are equilibrium averages of operators.
The conductivity is an even function of frequency, and it has only even moments  $\mu_{2k}$. For $k>0$, the moments are given by equilibrium averages
\bea
\mu_{2k} &=& {1\over \cV  } \langle j| \cL^{2k} |j) =  \left(L^{2k}[\Delta]\right)_{0,0} , \nonumber\\
&=& {1\over \cV  }  \Tr \rho \left[ j, \cL^{2k-1} j\right] .
\label{moments}
\eea
$L$ is the tridiagonal matrix given in Eq.~(\ref{Lmn}).
By taking the $(0,0)$ matrix elements of even powers $L[\Delta]$,  an algebraic recursive relation is obtained between the moments and recurrents
\bea
{\mu_2\over \mu_0}&=& \Delta_1^2  ,\nonumber\\
{\mu_4\over \mu_0}&=&  \Delta_1^2(\Delta_1^2+\Delta_2^2) , \nonumber\\
{\mu_6\over \mu_0} &=&\Delta_1^2\left( \Delta_1^4 + 2\Delta_1^2\Delta_2^2+\Delta_2^4+\Delta_2^2\Delta_3^3\right) , \nonumber\\
\vdots&=&\vdots ,
\label{mom-rec}
\eea
which can readily be inverted to obtain the lowest $k=1,2,\ldots k_{\rm max}$ recurrents from the lowest $k_{\rm max}+1$ moments
\bea
\Delta_1^2 &=& \mu_2\over \mu_0,\nonumber\\
\Delta_2^2 &=&{ \mu_4\over \mu_0 \Delta_1^2}-\Delta_1^2,\nonumber\\
\Delta_3^2&=&  {\mu_6 \over \mu_0 \Delta_1^2\Delta_2^2 }-{\Delta_1^4\over \Delta_2^2} - 2\Delta_1^2 - \Delta_2^2 ,\nonumber\\
\vdots&=&\vdots .
\label{Dn}
\eea

Note: a useful relation exists between the recurrents $\Delta_i, i\le n$ and the  normalization constants $N_n$ in Eq.~(\ref{Krylov})
\be
N_n =   \prod_{i=1}^{n} {1\over |\Delta_i| } .
\ee

\section{Variational Extrapolation of Recurrents}
\label{App:VER}
Any calculation of a {\em finite} set of recurrents $\Delta_n, n\le n_{\rm max}$ is  not  sufficient for
determining Eq.~(\ref{Gz1}). An infinite {\em extrapolation} of  $\{ \Delta_{ n'}\}_{n_{\rm max}+1}^{ \infty}$ is  required.

Several extrapolation schemes have been proposed.  The  Variational Extrapolation of Recurrents (VER)~\cite{Sandvik,lindner,khait} has been found to be reliable in certain cases.
VER  chooses a physically-motivated variational function  $\sigma^{\rm ver} (\omega; \{\alpha_i, i=1, \ldots  \} )$
with sufficiently many variational parameters. $ \alpha_i^{\rm ver} $ are  determined by a least-squares fit between the recurrents of $ \sigma^{\rm ver}$ and the
computed set. The conductivity is then approximated by,
\begin{widetext}
\be
\sigma_{\alpha\alpha}(\omega)  \approx      -  \Im  \cfrac{  \hbar \mu_0}{ \hbar\omega+ i\ve  -\cfrac{\Delta_1^2}{ \hbar\omega+ i\ve  -\cfrac{\Delta_2^2}{    \cfrac{  \ddots }{\hbar \omega+i\ve-\cfrac{|\Delta_{n_{\rm max}-1}|^2 }{\hbar\omega+ i\ve - |\Delta_{n_{\rm max}}|^2 T^{\rm ver}(\omega)  }}}}} ,
\ee  
\end{widetext}
where $T^{\rm ver}(\omega)$ is a complex termination function, which is ``borrowed'' from the fitted variational function $ \sigma^{\rm ver}(\omega)$.
The reliability of the VER procedure is in principle testable by finding convergence as  $n_{\rm max}$ is incrementally  increased.

 \section{Off-diagonal Green functions}
\label{App:OD}
To prove Eq.~(\ref{HallN}) we need to determine  $G_{n,m}(i\ve)= G'_{n,m}+ i G''_{n,m}$ in Eq.~(\ref{Gz}). Due to the tridiagonal properties of $L$ 
and the conditions $(G'+iG'')  L = L  (G'+iG'')  = I $, the following properties follow
\bea
&&G_{n,m}=G_{m,n} , \nonumber\\
&&G_{2i, 2j+1}  =     G'_{2i,2j+1} ={\rm real} , \nonumber\\
&&G_{2i,2j} = i G''_{2i,2j}  ={\rm imaginary} , \nonumber\\
&&G_{2i+1,2j+1} =   0 .
\label{Gn}
\eea
In particular, we see that for any  $|n\rangle$
\be
\Im G_{1,n}= \langle 1\Big| \left({1\over \cL}\right)'' \Big| n\rangle =0 .
\ee
Hence, one can write
\be
 \left({1\over \cL}\right)'' \cL  j^x  =0 ,
\ee
which proves Eq.~(\ref{delta-cl}).

The nonzero imaginary Green function can be written as
\bea
&& G''_{2k,0}   =  G''_{0,2k} =  R_k  G''_{0,0} , \nonumber\\
&&R_k  \equiv   \prod_{j=1}^{k} \left( - { \Delta_{2j-1}\over \Delta_{2j} } \right).
\label{Gn0}
\eea

By Eq.~(\ref{CF-sigma})
\be
\sigma_{xx} (0) = - \hbar \mu_0   G''_{0,0}  ,
\ee
where $\mu_0= (j^x|j^x)/\cV$.
Thus, by Eq. (\ref{Gn}), all the odd entries drop out of the sums  in Eq.~(\ref{HallN}), and $\sigma_{xx}^2$ factors out of the sums.
Therefore, the dissipative longitudinal conductivity is completely eliminated from the Hall coefficient formula, which  is left to depends solely on thermodynamical susceptibilities.

\bibliographystyle{unsrt}

\bibliography{HallN}

\begin{thebibliography}{10}

\bibitem{jones-zener}
H.~Jones.
\newblock H. Jones and C. Zener, 
\newblock { Proc. Roy. Soc.(London)}, 145, 268  (1934).

\bibitem{ziman}
J.~M. Ziman.
\newblock { Electrons and phonons: the theory of transport phenomena in
  solids}.
\newblock Oxford university press, (1960).

\bibitem{Ong}
N.~P. Ong.
\newblock Geometric interpretation of the weak-field hall conductivity in
  two-dimensional metals with arbitrary fermi surface.
\newblock { Phys. Rev. B}, 43,193- (1991).

\bibitem{Comm-tau}
{A}nisotropy of relaxation times on the {F}ermi surface effects the {H}all
  coefficient~\cite{ziman,Ong}. {N}everthelss, formula (\ref{HallC}) with its
  correction term is an equilibrium expression which includes the effects of
  scattering anisotropy. {S}ee discussion after {E}q.~(\ref{BS-tauk}).

\bibitem{Comm-Other}
{E}quilibrium expressions for { non-metallic } (i.e. zero resistivity)
  phases are the {S}treda formula and {C}hern number for the {H}all
  conductivity, which will be discussed in {S}ection \ref{Sec:Other}.

\bibitem{badmetals}
V.~J. Emery and S.~A. Kivelson.
\newblock Superconductivity in bad metals.
\newblock { Phys. Rev. Lett.}, 74,3253 (1995).

\bibitem{lindner}
N.~H. Lindner and A.~Auerbach.
\newblock Conductivity of hard core bosons: a paradigm of a bad metal.
\newblock { Phys. Rev. B}  81, 054512, 2010.

\bibitem{smith}
A.~W. Smith, T.~W. Clinton, C.~C. Tsuei, and C.~J. Lobb.
\newblock Sign reversal of the hall resistivity in amorphous mo 3 si.
\newblock { Phys. Rev. B}, 49, 12927  (1994).

\bibitem{kapitulnik}
X.~Zhang, Q.~Yang, A.~Palevski, and A.~Kapitulnik.
\newblock Superconductor-insulator transition in indium oxide thin films.
\newblock { Bulletin of the American Physical Society}, (2018).

\bibitem{hagen}
S.~J. Hagen, C.~J. Lobb, R.~L. Greene, M.~G. Forrester, and J.~H. Kang.
\newblock Anomalous hall effect in superconductors near their critical
  temperatures.
\newblock { Phys. Rev. B}, 41, 11630 (1990).

\bibitem{Taillefer}
S.~Badoux, W.~Tabis, F.~Lalibert{\'e}, G.~Grissonnanche, B.~Vignolle,
  D.~Vignolles, J.~B{\'e}ard, D.~A. Bonn, W.~N. Hardy, R.~Liang,
  N.~Doiron-Leyraud, L. Taillefer, and C.  Proust.
\newblock Change of carrier density at the pseudogap critical point of a
  cuprate superconductor.
\newblock { Nature}, 531, 210 (2016).

\bibitem{Lior}
L.~Klein, J.S. Dodge, C.H. Ahn, J.W. Reiner, L~Mieville, T.H. Geballe, M.R.
  Beasley, and A.~Kapitulnik.
\newblock Transport and magnetization in the badly metallic itinerant
  ferromagnet.
\newblock { Journal of Physics: Condensed Matter}, 8, 10111 (1996).

\bibitem{Assa-PRL}
Assa Auerbach.
\newblock Hall number of strongly correlated metals.
\newblock { Phys. Rev. Lett.}, 121, 066601, (2018).

\bibitem{domb}
C.~Domb.
\newblock { Phase transitions and critical phenomena, Vol 3,}, volume~19.
\newblock Elsevier, (2000).

\bibitem{DMRG}
U. Schollwock.
\newblock The density-matrix renormalization group in the age of matrix product
  states.
\newblock { Annals of Physics}, 326(1), 96 (2011).
\newblock January 2011 Special Issue.

\bibitem{prokofiev}
N.~Prokof'ev and B.~Svistunov.
\newblock Worm algorithms for classical statistical models.
\newblock { Phys. Rev. Lett.}, 87,160601  (2001).

\bibitem{assaad}
F.~F. Assaad.
\newblock Phase diagram of the half-filled two-dimensional {SU(N)}
  {H}ubbard-{H}eisenberg model: {A} quantum {M}onte {C}arlo study.
\newblock { Phys. Rev. B}, 71, 075103 (2005).

\bibitem{Comm-RT}
{R}eal-time dissipative response functions suffer from {S}uzuki-{T}rotter
  errors at long times~\cite{White}, ill-posed analytical continuation at
  finite temperatures~\cite{MaxEnt,snir-QMC}, or extrapolation of moments to
  infinite order~\cite{viswanath,lindner,khait}. see appendix (\ref{App:VER})
  for a review of such an extrapolation.

\bibitem{White}
S.R. White and A.E. Feiguin.
\newblock Real-time evolution using the density matrix renormalization group.
\newblock { Phys. Rev. Lett.}, 93,076401, (2004).

\bibitem{MaxEnt}
M.~Jarrell and J.~E. Gubernatis.
\newblock Bayesian inference and the analytic continuation of imaginary-time
  quantum monte carlo data.
\newblock { Physics Reports}, 269, 133 (1996).

\bibitem{snir-QMC}
S.~Gazit, D.~Podolsky, A.~Auerbach, and D.P. Arovas.
\newblock Dynamics and conductivity near quantum criticality.
\newblock { Phys. Rev. B}, 88, 235108 (2013).

\bibitem{behnia-review}
K.~Behnia and H.~Aubin.
\newblock Nernst effect in metals and superconductors: a review of concepts and
  experiments.
\newblock { Reports on Progress in Physics}, 79(4),046502 (2016).

\bibitem{shastry-sumrules}
B.~Sriram Shastry.
\newblock Sum rule for thermal conductivity and dynamical thermal transport
  coefficients in condensed matter.
\newblock { Phys. Rev. B}, 73, 085117 (2006).

\bibitem{Bogoliubov}
N.~N. Bogoliubov.
\newblock { Dubna Report}, 1962.

\bibitem{MerminWagner}
N.~D. Mermin and H.~Wagner.
\newblock Absence of ferromagnetism or antiferromagnetism in one- or
  two-dimensional isotropic heisenberg models.
\newblock { Phys. Rev. Lett.}, 17, 1133  (1966).

\bibitem{Mori}
H.~Mori.
\newblock Transport, collective motion, and brownian motion.
\newblock { Progress of Theoretical Physics}, 33, 423  (1965).

\bibitem{forster}
D.~Forster.
\newblock { Hydrodynamic fluctuations, broken symmetry, and correlation
  functions, Ch. V}.
\newblock CRC Press, (2018).

\bibitem{zwanzig}
R.~Zwanzig.
\newblock { Nonequilibrium statistical mechanics}.
\newblock Oxford University Press, (2001).

\bibitem{wolfle}
W.~G{\"o}tze and P.~W{\"o}lfle.
\newblock Homogeneous dynamical conductivity of simple metals.
\newblock { Phys. Rev. B}, 6, 1226, (1972).

\bibitem{MHLee}
M.~Howard Lee, J.~Hong, and J.~Florencio~Jr.
\newblock Method of recurrence relations and applications to many-body systems.
\newblock { Physica Scripta}, 1987, 498, (1987).

\bibitem{viswanath}
V.~S. Viswanath and G.~M{\"u}ller.
\newblock { The Recursion Method: Application to Many Body Dynamics},
  volume~23.
\newblock Springer Science \& Business Media, (1994).

\bibitem{Sandvik}
O.~A. Starykh, A.~W. Sandvik, and R.~R.~P. Singh.
\newblock Dynamics of the spin-heisenberg chain at intermediate temperatures.
\newblock { Phys. Rev. B}, 55, 14953, (1997).

\bibitem{khait}
I.~Khait, S.~Gazit, N.~Y. Yao, and A.~Auerbach.
\newblock Spin transport of weakly disordered heisenberg chain at infinite
  temperature.
\newblock { Phys. Rev. B}, 93, 224205 (2016).

\bibitem{Streda}
P.~Streda and L.~Smrcka.
\newblock Thermodynamic derivation of the hall current and the thermopower in
  quantising magnetic field.
\newblock { Journal of Physics C: Solid State Physics}, 16, L895, (1983).

\bibitem{McD}
A.~H. MacDonald and P.~Steda.
\newblock Quantized hall effect and edge currents.
\newblock { Phys. Rev. B}, 29,1616,  (1984).

\bibitem{TKNN}
D.~J. Thouless, M.~Kohmoto, M.~P. Nightingale, and M.~den Nijs.
\newblock Quantized hall conductance in a two-dimensional periodic potential.
\newblock { Phys. Rev. Lett.}, 49, 405 (1982).

\bibitem{yosi}
J.~E. Avron and R.~Seiler.
\newblock Quantization of the hall conductance for general, multiparticle
  schr{\"o}dinger hamiltonians.
\newblock { Phys. Rev.  Lett. }, 54, 259 (1985).

\bibitem{huber}
S.~D. Huber and N.~H. Lindner.
\newblock Topological transitions for lattice bosons in a magnetic field.
\newblock { Proceedings of the National Academy of Sciences},
  108, 19925 (2011).

\bibitem{Chern-pump-theory}
A.~Dauphin and N.~Goldman.
\newblock Extracting the chern number from the dynamics of a fermi gas:
  Implementing a quantum hall bar for cold atoms.
\newblock { Phys. Rev. Lett.}, 111,135302 (2013).

\bibitem{prelov-ladder}
P.~Prelov{\v{s}}ek, M.~Long, T.~Marke{\v{z}}, and X.~Zotos.
\newblock Hall constant of strongly correlated electrons on a ladder.
\newblock { Phys. Rev. Lett.}, 83,  2785 (1999).

\bibitem{Chern-pump-exp}
M.~Aidelsburger, M.~Lohse, C.~Schweizer, M.~Atala, J.~T. Barreiro,
  S.~Nascimbene, N.~R. Cooper, I.~Bloch, and N.~Goldman.
\newblock Measuring the chern number of hofstadter bands with ultracold bosonic
  atoms.
\newblock { Nature Physics}, 11, 162, 2015.

\bibitem{Cooper}
N.~R. Cooper, B.~I. Halperin, and I.~M. Ruzin.
\newblock Thermoelectric response of an interacting two-dimensional electron
  gas in a quantizing magnetic field.
\newblock { Phys. Rev. B}, 55,2344  (1997).

\bibitem{Comm-derivation}
{T}he derivation below is simpler and more general than in
  {R}ef.~\cite{Assa-PRL}. {T}he primary difference is the use of polarizations
  operators, which mutually commute, and are independent of magnetic field.

\bibitem{Hardy}
R.~J. Hardy.
\newblock Energy-flux operator for a lattice.
\newblock { Phys. Rev.}, 132,168 (1963). 

\bibitem{SW}
J.~R. Schrieffer and P.~A. Wolff.
\newblock Relation between the anderson and kondo hamiltonians.
\newblock { Phys. Rev.}, 149, 491 (1966).

\bibitem{IEQM}
A.~Auerbach.
\newblock { Interacting electrons and quantum magnetism}.
\newblock Springer Science \& Business Media (2012).

\bibitem{CORE-marvin}
C.~J. Morningstar and M.~Weinstein.
\newblock Contractor renormalization group technology and exact hamiltonian
  real-space renormalization group transformations.
\newblock { Phys. Rev. D}, 54, 4131 (1996).

\bibitem{CORE-altman}
E.~Altman and A.~Auerbach.
\newblock Plaquette boson-fermion model of cuprates.
\newblock { Phys. Rev. B}, 65, 104508 (2002).

\bibitem{p6}
S.~Capponi, V.R. Chandra, A.~Auerbach, and M.~Weinstein.
\newblock p 6 chiral resonating valence bonds in the kagome antiferromagnet.
\newblock { Phys. Rev. B}, 87, 161118 (2013).

\bibitem{behnia-book}
K.~Behnia.
\newblock { Fundamentals of thermoelectricity}.
\newblock OUP Oxford, 2015.

\bibitem{BHM-Fisher}
M.~P.~A. Fisher, P.~B. Weichman, G.~Grinstein, and Daniel~S. Fisher.
\newblock Boson localization and the superfluid-insulator transition.
\newblock { Phys. Rev. B}, 40, 546 (1989).

\bibitem{SSS}
B.~S. Shastry, B.~I. Shraiman, and R.~R.~P. Singh.
\newblock Faraday rotation and the hall constant in strongly correlated fermi
  systems.
\newblock { Phys. Rev.  Lett. }, 70, 2004 (1993).

\bibitem{Com3}
The minimal coupling {$\nabla\varphi(\bx) +{q\over c} \bA$} is consistent with
  the sign of the density-phase commutator.

\bibitem{Girvin}
M.~Jonson and S.M. Girvin.
\newblock Thermoelectric effect in a weakly disordered inversion layer subject
  to a quantizing magnetic field.
\newblock { Phys. Rev. B}, 29,1939 (1984).

\bibitem{Comm-Streda}
A weaker condition for reversing the order of limits, is local
  incompressibility~\cite{MD}. that is to say, the charge and flux densities
  (both coarse grained quantities on some microscopic correlation length)
  should be ``locally locked'' $\langle \rho(\bx) \rangle \propto \langle
  b(\bx)\rangle $, and similarly, $\langle s(\bx) \rangle \propto \langle
  b(\bx)\rangle $.

\bibitem{lindnerHall}
N.~Lindner, A.~Auerbach, and D.~P. Arovas.
\newblock Vortex dynamics and hall conductivity of hard-core bosons.
\newblock { Phys. Rev. B}, 82,134510 (2010).

\bibitem{berg}
E.~Berg, S.~D. Huber, and N.~H. Lindner.
\newblock Sign reversal of the hall response in a crystalline superconductor.
\newblock { Phys. Rev. B}, 91, 024507 (2015).

\bibitem{Assa-QH}
A. Auerbach.
\newblock Comparison of tunneling rates of fractional charges and electrons
  across a quantum hall strip.
\newblock { Phys. Rev.  Lett. }, 80, 817 (1998).

\bibitem{Ilia}
I. Khait and A. Auerbach.
\newblock {H}all number of the {t-J} model.
\newblock { Manuscript in preparation}.

\bibitem{hewson}
A.~C. Hewson.
\newblock { The Kondo problem to heavy fermions}, volume~2.
\newblock Cambridge university press, (1997).

\bibitem{ari-weyl}
X.~Wan, A.~Turner, A.~Vishwanath, and S.~Y.~Savrasov.
\newblock Topological semimetal and fermi-arc surface states in the electronic
  structure of pyrochlore iridates.
\newblock { Phys. Rev. B}, 83(20),205101 (2011).

\bibitem{Fuku}
H.~Fukuyama.
\newblock Effects of intervalley impurity scattering on the non-metallic
  behavior in two-dimensional systems.
\newblock { Journal of the Physical Society of Japan}, 49(2),649 (1980).

\bibitem{Altshuler}
B.~L. Altshuler, D.~Khmel'nitzkii, A.~I. Larkin, and P.~A. Lee.
\newblock Magnetoresistance and hall effect in a disordered two-dimensional
  electron gas.
\newblock { Phys. Rev. B}, 22, 5142 (1980).

\bibitem{Boris}
B.~Shapiro and E.~Abrahams.
\newblock Scaling theory of the hall effect in disordered electronic systems.
\newblock { Phys. Rev. B}, 24, 4025 (1981).

\bibitem{AS}
E. Shimshoni and A. Auerbach.
\newblock Quantized hall insulator: Transverse and longitudinal transport.
\newblock { Phys. Rev. B}, 55, 9817 (1997).

\bibitem{MD}
A.~H. MacDonald and P.~Steda.
\newblock Quantized hall effect and edge currents.
\newblock { Phys. Rev. B}, 29, 1616 (1984).

\end{thebibliography}

 \end{document}